# Tunable Two-Dimensional Group-III Metal Alloys

Siavash Rajabpour[1,2], Alexander Vera[2,3], Wen He[4,5], Benjamin N. Katz[2,6], Roland J. Koch[7], Margaux Lassaunière[8,9], Xuegang Chen[10], Cequn Li[2,6], Katharina Nisi[8,11], Hesham El-Sherif[12], Maxwell T. Wetherington[3,13], Chengye Dong[2,14], Aaron Bostwick[7], Chris Jozwiak[7], Adri C.T. van Duin,[1,2,3,13,14,15,16,17], Nabil Bassim[12,18], Jun Zhu[2,6], Gwo-Ching Wang[10], Ursula Wurstbauer[8,9], Eli Rotenberg[7], Vincent Crespi[2,3,6,13,14,16], Su Ying Quek[4,5,19,20], Joshua A. Robinson.[2,3,13,14]∗

1. Department of Chemical Engineering, The Pennsylvania State University, University Park, PA, USA
2. Center for 2-Dimensional and Layered Materials, The Pennsylvania State University, University Park, PA, USA
3. Department of Materials Science and Engineering, The Pennsylvania State University, University Park, PA, USA
4. Department of Materials Science and Engineering, National University of Singapore, 9 Engineering Drive, Singapore, Singapore
5. Centre for Advanced 2D Materials, National University of Singapore, 6 Science Drive 2, Singapore, Singapore
6. Department of Physics, The Pennsylvania State University, University Park, PA, USA
7. Advanced Light Source, Lawrence Berkeley National Laboratory, Berkeley, California, USA
8. Institute of Physics, University of Münster, Münster, Germany
9. Center for Soft Nanoscience, University of Münster, Münster, Germany
10. Department of Physics, Applied Physics and Astronomy, Rensselaer Polytechnic Institute, Troy, NY, USA
11. Physics Department, Technical University of Munich, Garching, Germany
12. Department of Materials Science and Engineering, McMaster University, Hamilton, Ontario, Canada
13. Materials Research Institute, The Pennsylvania State University, University Park, PA, USA
14. 2-Dimensional Crystal Consortium, The Pennsylvania State University, University Park, PA, USA.
15. Department of Mechanical Engineering, The Pennsylvania State University, University Park, PA, USA
16. Department of Chemistry, The Pennsylvania State University, University Park, PA, USA.
17. Department of Engineering Science and Mechanics, The Pennsylvania State University, University Park, PA, USA
18. Canadian Centre for Electron Microscopy, Hamilton, Ontario, Canada
19. Department of Physics, National University of Singapore, Singapore, Singapore
20. NUS Graduate School Integrative Sciences and Engineering Programme, National University of Singapore, Singapore 117456

* jrobinson@psu.edu

**Abstract**

Chemically stable quantum-confined 2D metals are of interest in next-generation nanoscale quantum devices. Bottom-up design and synthesis of such metals could enable the creation of materials with tailored, on-demand, electronic and optical properties for applications that utilize tunable plasmonic coupling, optical non-linearity, epsilon-near-zero behavior, or wavelength-specific light trapping. In this work, we demonstrate that the electronic, superconducting and optical properties of air-stable two-dimensional metals can be controllably tuned by the formation of alloys. Environmentally robust large-area two-dimensional $In_xGa_{1-x}$ alloys are synthesized by Confinement Heteroepitaxy (CHet). Near-complete solid solubility is achieved with no evidence of phase segregation, and the composition is tunable over the full range of $x$ by changing the relative elemental composition of the precursor. The optical and electronic properties directly correlate with alloy composition, wherein the dielectric function, band structure, superconductivity, and charge transfer from the metal to graphene are all controlled by the indium/gallium ratio in the 2D metal layer.

**Introduction**

Two-dimensional mono-elemental materials exhibit a wide range of optical, electronic, and superconducting properties that diverge from those of their bulk counterparts[1–4]. For example, few-layer indium on Si(111)[5] and gallium on GaN(0001)[6] show increased superconducting transition temperatures, potentially facilitating their use in nanoscale quantum devices. Ultrathin gold exhibits strongly tunable plasmons with resonance peaks between 1.5 and 5 μm, suggesting foundational applications in plasmonic metasurfaces[7]. These property enhancements have benefited from refinements in synthesis by evaporation[8], liquid exfoliation[9,10], and molecular beam epitaxy[2]. However, the ability to intercalate various metal species into high-energy interfaces[11] suggests an alternative synthesis route towards stabilizing two-dimensional metals not previously pursued. This route, dubbed "confinement heteroepitaxy" (CHet)[12], enables the formation of environmentally robust 2D metals that are readily studied by a wide variety of ex-situ techniques. In CHet, metal atoms are intercalated into the gallery between epitaxial graphene (EG) and SiC(0001) under near-atmospheric pressure in argon environment, to form a large-area 2D crystals epitaxial to the SiC lattice. Similar to 2D metal systems grown by molecular beam epitaxy under ultra-high vacuum[2,4], CHet-based 2D metals exhibit clean, abrupt interfaces and properties distinct from their bulk counterparts. For CHet-derived systems, these include zero-crossings in the frequency-dependent dielectric constant[13], extraordinary room-temperature nonlinear susceptibilities $\chi^{(2)}$ >10 nm/V[14], and a 4-fold increase in the superconducting transition temperature over that of the most stable bulk phase[12]. When considered in the context of the environmental robustness of CHet-based 2D metals, these properties strongly suggest the potential utility within multicomponent nanoscale photonic and quantum devices.

The prospect of tuning 2D metal properties via doping and alloying promises to further expand the materials frontier and technological impact of these two-dimensional metals. Early experiments toward this end, such as surface alloy phases on bulk metals[15–23] and the intercalation of $Sn_{1-x}Ge_x$[24] at the EG/SiC(0001) interface motivate exploration of how 2D metal alloy composition enables tuning of the inter-band contributions to dielectric properties, alloy-dependent Fermi surface geometry, and other aspects of optical and electronic response. Here, we report the controlled synthesis and characterization of two-dimensional $In_xGa_{1-x}$ alloys via CHet, using high-purity In and Ga metal precursors. We find the 2D metal alloy composition is readily tunable by changing the elemental concentration in the precursor, confirmed through high-resolution x-ray photoelectron spectroscopy (XPS) and auger electron spectroscopy (AES) mapping. Cross-sectional scanning transmission electron microscopy (STEM) and azimuthal reflection high-energy electron diffraction (ARHEED) confirm the metals are epitaxial to the SiC substrate, and electron energy loss spectroscopy (EELS) demonstrates metal confinement between graphene and SiC. First-principles density functional theory (DFT) predicts and STEM verifies that the layers are randomly mixed (approaching an ideal alloy) with a nominal thickness of two atoms. The optical and electronic properties directly correlate with alloy composition, where the dielectric response, electronic band structure, superconductivity, and charge transfer from metal to graphene all vary in controlled ways with respect to the indium concentration in the 2D layer.

**Results and Discussions**

Atomically thin $In_xGa_{1-x}$ alloys are realized through confinement heteroepitaxy (CHet) (see Methods), where group-III metal atoms are thermally evaporated at near-atmospheric pressures and then stabilized as 2D metal alloys by intercalation into a defect-engineered EG/SiC interface. Tuning of the 2D $In_xGa_{1-x}$ alloy composition is achieved by simply adjusting the alloy precursor composition, and thus metal atom flux, from 0 to 1 mole fraction indium. X-ray photoelectron spectroscopy (**Figure 1a,b**) provides evidence that the metal is successfully intercalated at the EG/SiC interface, where the splitting of the C1s peak signifies passivation of the buffer layer by the intercalant[12,25] (**See SI, Figure S1**). Furthermore, XPS confirms both In and Ga at the EG/SiC interface, where a slight shift in the In 3d and Ga 2p spectra occurs in the 2D $In_xGa_{1-x}$ alloy compared to pure 2D In and 2D Ga, signifying the variation in chemical bonding environment introduced by the presence of a second element. Finally, XPS reveals that the correlation between the composition of the $In_xGa_{1-x}$ precursor and that of the 2D-$In_xGa_{1-x}$ alloy is nearly linear for the compositions examined (**Figure 1c**), indicating that specific 2D alloy compositions across the full range of $0 < x < 1$ can likely be achieved within a few percent by tuning the precursor composition. The near-linear relation between precursor and final alloy compositions is remarkable, and reflects an interplay of kinetic and equilibrium aspects of metal intercalation in the experimental geometry.

Using CHet, the $In_xGa_{1-x}$ alloys are uniformly distributed with no apparent lateral segregation. As evidenced by AES mapping (**Figure 1d**), the uniform spectral intensity of Ga and In as a function of position in the 2D $In_{0.5}Ga_{0.5}$ alloy (other alloys are similar) demonstrates that both metals intercalate with similar overall efficiencies and exhibit an essentially uniform mixing upon intercalation. This suggests that the group-III elements react similarly with graphene defects formed during CHet, following similar transport kinetics within the intercalation gallery, this is likely due to their similarity in elemental properties and may not be the case for highly disparate elements such as mixtures of group-III and refractory d-block elements. We note that the diagonal stripes in the AES maps are associated with regions of multi-layer EG that often occur at the step edges in SiC that separate the (0001) terrace regions. Further evidence of this can be found in the C and Si maps where step edge regions exhibit an increased carbon and Si signals **(Figure S2)**. Despite this, Ga and In atoms are continuous and uniform across the micron-scale terraces, with minimal oxygen present. Regions with increased oxygen are encountered, generally where the EG is scratched or EG layer formation is incomplete. In those cases, we find severe oxidation of the 2D metals, validating the critical role of the graphene overlayer to maintaining environmental stability **(Figure S3)**.

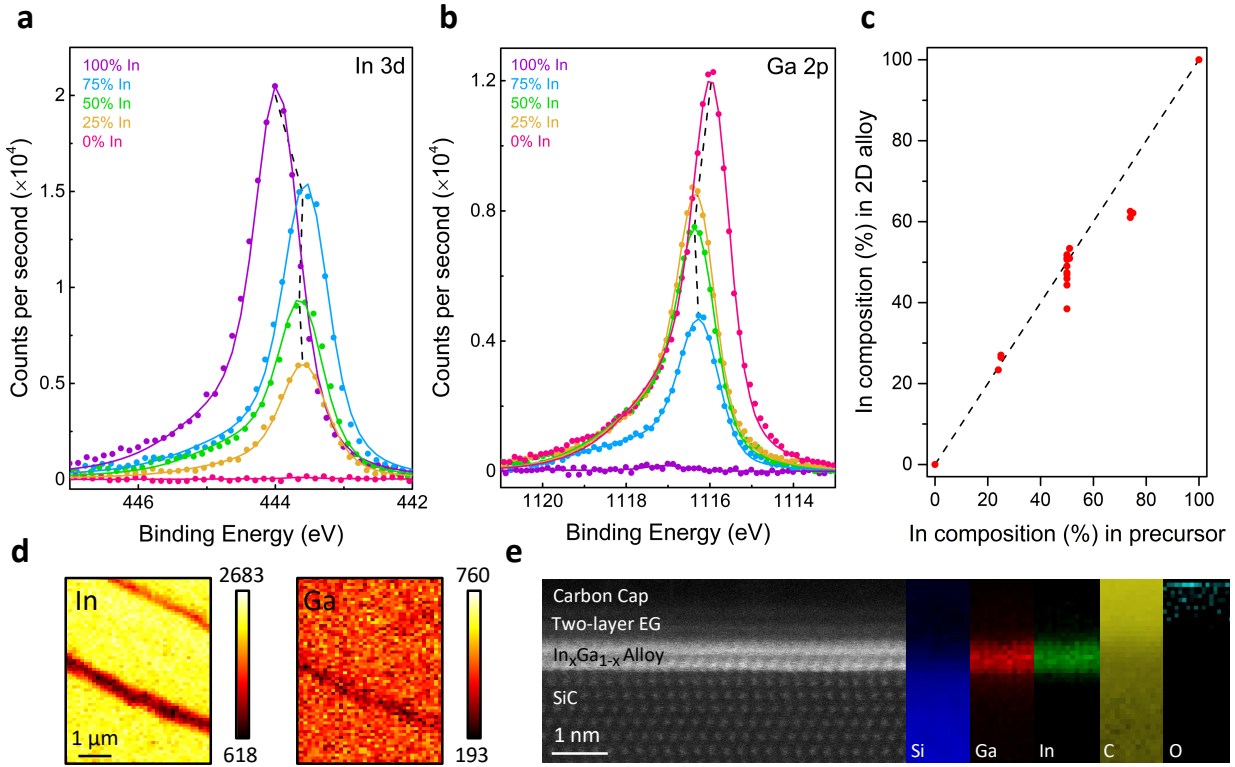

**Figure 1. Compositional and structural characterization of CHet-grown 2D-In$_x$Ga$_{1-x}$ alloys. (a,b)** XPS analyses verify the co-existence of metallic Ga and In in the alloys. Ga 2p and In 3d peaks are identified close to 1116 and 444 eV, which are the metallic peak positions for pure 2D-Ga and 2D-In, respectively. Slight shifts in Ga 2p and In 3d peak positions for the 2D-In$_x$Ga$_{1-x}$ alloys, compared to the pure 2D-Ga and 2D-In, correspond to different neighbors' chemistries introduced by the second element in the alloys. **(c)** Tuning the composition of the 2D-In$_x$Ga$_{1-x}$ alloys is possible by adjusting the precursor composition, with close correspondence between alloy and precursor compositions. **(d)** Ga and In AES maps of the 2D-In$_{0.5}$Ga$_{0.5}$ alloy showing uniform distribution of Ga and In across the mapped regions. The diagonal regions of low signal correspond to steps in the SiC which can contain additional EG layers that attenuate the In and Ga signals. **(e)** Cross-sectional STEM and accompanying EELS mapping showing two layers of uniformly dispersed 2D-In$_{0.5}$Ga$_{0.5}$ alloy between EG and SiC, where EELS verifies oxygen is not present.

Two dimensional In$_x$Ga$_{1-x}$ alloys are bilayer epitaxial films, confined at the EG/SiC(0001) interface. High angle annular dark-field (HAADF) STEM imaging (**Figure 1e**) demonstrates uniform contrast in the alloy layer, indicating a uniform dispersion of Ga and In atoms over the regions examined. Electron energy loss spectroscopy (EELS) mapping further verifies the uniform distribution of Ga and In (**Figure S4**), where the 2D metal layer exhibits co-located Ga-L$_{2,3}$ and In-M$_{4,5}$ EELS signals without oxygen. Further quantification of the In and Ga composition at the atomic scale using core-loss EELS, applied to the In$_{0.5}$Ga$_{0.5}$ alloy, further verifies the layers are uniform in composition. Beyond a uniform dispersion of In and Ga, STEM imaging also indicates the 2D-In$_x$Ga$_{1-x}$ alloys are crystalline (**Figure 1e**). This is verified by RHEED (**Figure 2**) of EG/2D-In$_{0.5}$Ga$_{0.5}$/SiC at two azimuthal angles rotated 30° to each other. The patterns (**Figure 2a, b**) exhibit vertical streaks perpendicular to the surface, indicating a smooth surface. Non-vertical streaks and spots near the center streak (labeled 00) are also observed from the SiC substrate alone, indicating they are not related to the EG/2D-In$_{0.5}$Ga$_{0.5}$. Plots in **Figure 2c, d** of the intensity versus momentum transfer parallel to the surface ($\mathbf{k}_\parallel$) from the boxed regions in **Figure 2a,b** (chosen to avoid the substrate signal) demonstrate an average spacing ($\Delta \mathbf{k}_\parallel$) between adjacent peaks of 3.04 ± 0.14 Å$^{-1}$. $\Delta \mathbf{k}_\parallel$ is related to the real space lattice constant by $a = 2\pi/[(\sqrt{3}/2) \Delta \mathbf{k}_\parallel]$; here $a$ = 2.38 ± 0.11 Å, which is consistent with graphene $a$ = 2.46 Å. Therefore, the measured $\Delta \mathbf{k}_\parallel$ of 3.04 ± 0.14 Å$^{-1}$ is assigned as reciprocal space vector **G**(01). The peaks observed in **Figure 2c** are assigned as $0\bar{2}$, $0\bar{1}$, 00, 01, and 02 peaks of graphene. This assignment is further verified when $\Delta \mathbf{k}_\parallel$ is measured at a 30° azimuthal angle from **Figure 2d**. The average $\Delta \mathbf{k}_\parallel$ = 5.23 ± 0.03 Å$^{-1}$, close to the length of the reciprocal space vector **G**($1\bar{2}$) = 5.108 Å$^{-1}$ for graphene.

Two dimensional In$_x$Ga$_{1-x}$ is matched in lattice constant and symmetry to the SiC substrate. This epitaxial relation is evident when considering a 2D reciprocal space map of EG/2D-In$_{0.5}$Ga$_{0.5}$/SiC (**Figure 2e, f**). A grid is constructed using the unit vector **a*** along G(01), and unit vector **b*** along G(10), 60° rotated with the same magnitude. Most a*/b* intersections coincide with diffraction spots from graphene. However, some diffraction spots between the 1$^{st}$ order spots of graphene, labelled by dark blue arrows, are not at the intersections of the graphene grid. RHEED patterns and symmetry from a 2D map of the SiC substrate (**Figure S5a**) measure a reciprocal space vector **G**(01) = **a**$_s$* = 2.36 ± 0.03 Å$^{-1}$ and **b**$_s$* = G(10), with a$_s$* = b$_s$*. The corresponding real space lattice constant $a$ = 3.07 ± 0.04 Å matches bulk SiC. Like the black grid from graphene on the EG/2D-In$_{0.5}$Ga$_{0.5}$/SiC 2D reciprocal space map in **Figure 2e**, a dark blue grid in **Figure 2f** consisting of **a**$_s$* and **b**$_s$* superposed on the experimental 2D map of the same EG/2D-In$_{0.5}$Ga$_{0.5}$/SiC 2D map coincides with the six 1$^{st}$ order spots located between the 1$^{st}$ order graphene spots. The unit vectors **a**$_s$* and **b**$_s$* of SiC are rotated 30° relative to **a*** and **b*** of graphene, consistent with EG/2D-In/SiC observed by low-energy electron diffraction[12]. These results indicate that there are no other regular reciprocal space lattices besides graphene and SiC. The combination of STEM and 2D maps from ARHEED thus provides evidence that the 2D metal alloys are lattice matched to the SiC substrate, just as are the mono-elemental end-members[12].

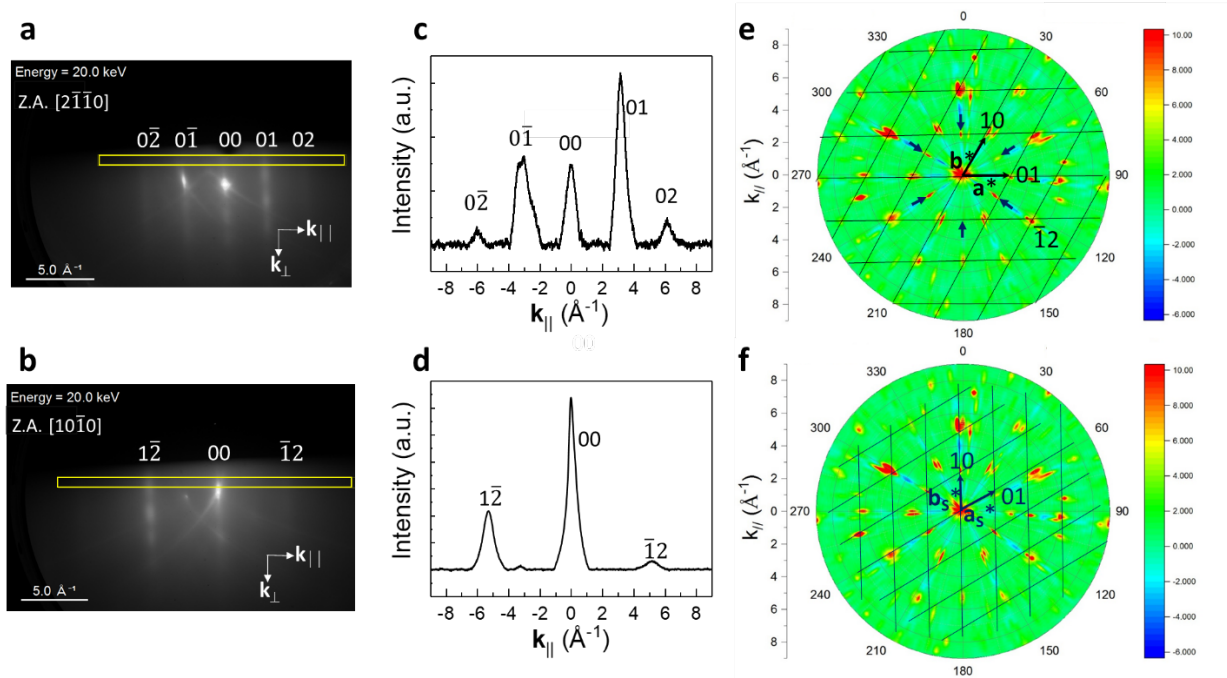

**Figure 2.** Reflection high-energy electron diffraction (RHEED) patterns taken from EG/2D-In$_{0.5}$Ga$_{0.5}$/SiC at 20 keV electron energy along the zone axis (Z.A.) **(a)** [2$\bar{1}\bar{1}$0] and **(b)** [10$\bar{1}$0]. Each visible diffraction streak is labeled by its Miller index. The two arrows perpendicular to each other represent the momentum transfers $\mathbf{k}_\parallel$ and $\mathbf{k}_\perp$ parallel and perpendicular to the surface. **(c)** and **(d)** Intensity profiles scanned along the $\mathbf{k}_\parallel$ direction within the yellow boxes ($|\mathbf{k}_\perp| = 4.5$ Å$^{-1}$) in the corresponding RHEED patterns shown in **(a)** and **(b)** after smooth background intensity subtraction. **(e)** and **(f)** Same experimentally constructed ARHEED 2D reciprocal space map of EG/2D-In$_{0.5}$Ga$_{0.5}$/SiC sliced parallel to surface at $|\mathbf{k}_\perp| = 8.5$ Å$^{-1}$. Intersections of the black grid in **(e)** correspond to the graphene lattice based on unit vectors $\mathbf{a}^* = \mathbf{b}^*$. Intersections of dark blue grid in **(f)** is the SiC lattice based on unit vectors $\mathbf{a}_s^* = \mathbf{b}_s^*$.

Only modest differences in total energies (< 30 meV per metal atom) exist for a wide range of 2D-In$_x$Ga$_{1-x}$ structures, suggesting no preference in elemental arrangement in the 2D metal alloy. This is supported by DFT calculations of twenty-five structures (**Figure S6, Table S1**) covering seven distinct alloy compositions for monolayer, bilayer, and trilayer systems – one configuration each for monolayer systems at 0, 8, 92, and 100% indium; two configurations each for monolayers at 25, 50, and 75% indium; four configurations each for bilayers at 25, 50, and 75% indium; pure bilayers of Ga and In, and a pure Ga trilayer. There is a weak preference towards layer segregation (~0.1 eV per metal atom or less) for indium at the bottom layer at low and high indium concentrations and gallium in the bottom layer at intermediate compositions; the weakness of this preference, considering the high temperature of intercalation, is consistent with the well-mixed layers observed experimentally. **Figure 3** displays a plot of the most favorable alloy composition at given chemical potentials for indium and gallium. The bilayer is relatively more favorable for gallium than the monolayer, as evidenced by the general leftwards shift in the bilayer phase boundaries relative to the monolayer regions. The slices through this plot (**Figure 3b and 3c**) show how close competing concentrations can be even across a wide range of chemical potential; overall, the results indicate a lack of line phases, a preference for well-mixed alloys, and some potential for compositional fluctuations where several compositions are closely competing at a given chemical potential.

A simple nearest-neighbor cluster expansion with distinct pair-wise interactions between In-In, Ga-Ga, and In-Ga within or across lower and upper layers and an on-site energy can reproduce the first-principles total energies of this set of alloy configurations to within a few hundredths of an eV/metal atom (these errors and the associated fitting parameters are given in **Tables S1** and **S2** respectively). Two additional extremal (1/12 and 11/12) In concentrations were included in the nearest-neighbor model to capture the effects of isolated minority atoms, but are omitted from the phase diagram of **Figure 3a**. Neither gallium nor indium naturally prefer the SiC-induced lattice spacing of 2.97 Å. Comparison to bulk inter-atomic spacings in pure metals suggests that gallium is under modest tensile strain[12] and indium is under large compressive strain when epitaxial to SiC (while indium does not have a single nearest-neighbor distance in bulk, even the shortest distance at 3.24 Å[26] is much longer than the SiC spacing), consistent with a net preference for gallium intercalation.

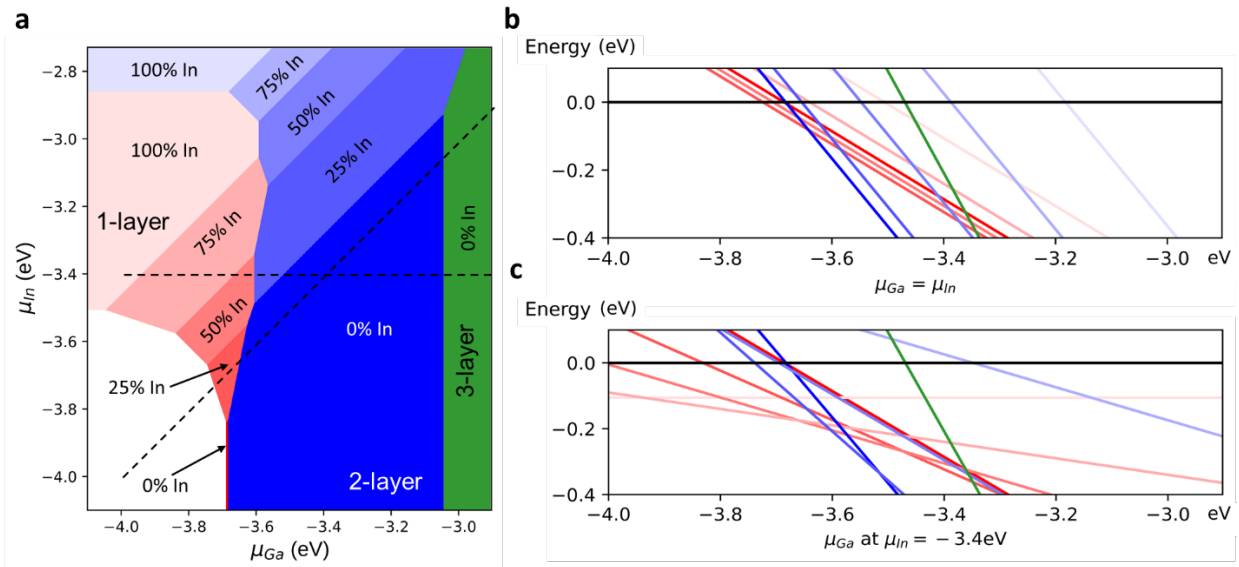

**Figure 3. (a)** Alloy stability as a function of Ga and In chemical potentials and layer thickness, based on first-principles Gibbs free energies including configurational entropy of ideal mixing at the growth temperature of 800 °C. **(b,c)** Two cuts along the dashed lines show competing phases.

Having established the structure and solubility properties of our as-prepared 2D $In_xGa_{1-x}$ alloys, we now systematically investigate the electronic and optical properties of the alloys as a function of composition. Angle-resolved photoemission spectroscopy (ARPES) was performed on the 2D $In_xGa_{1-x}$ alloys. **Figure 4a** and **4b** show the measured band dispersion and Fermi surface for the 2D $In_{0.75}Ga_{0.25}$ alloy. The most intense (bright blue) points of **Figure 4b** correspond to the Dirac point of graphene. More extended Fermi surface structures are also seen; these arise from bands of the 2D metal alloy. Superimposed on the experimental ARPES data in **Figure 4a** is a DFT band structure in the virtual crystal approximation[27] (VCA) computed for a 2D $In_{0.75}Ga_{0.25}$ alloy on SiC, without the overlying bilayer graphene (see Methods and **Figure S7** for methodology and see **Figure S8** for additional comparisons at different compositions). In the virtual crystal approximation, the potentials of In and Ga are mixed to capture their average effect at each atomic site in the 2D metal. The Fermi level within the VCA is shifted to match the ARPES bandstructure (which reasonably reproduces the measured Fermi surface geometry), as charge transfer to graphene from the metal (absent in the calculation) is significant. The computed band structure for the occupied states agrees well with that from ARPES across at least 2 eV from the Fermi level. This good agreement between the VCA calculations and experiment further confirms the near-complete solid solubility of In and Ga in $In_xGa_{1-x}$ alloys. We note that the VCA band structure for $In_{0.5}Ga_{0.5}$ is comparable to the band structure computed for two explicit (i.e. non-VCA) low-energy alloy structures of the same composition (shown in **Figure S9)**, where the band structure for the explicit alloy calculations has been unfolded into the effective primitive Brillouin zone (**Figure S9**). Similar to 2D Ga[12], the comparison between theory and experiment indicates that the bilayer graphene does not hybridize significantly with the metallic states close to the Fermi level. In both the ARPES data and the VCA calculations, the band dispersion in the 2D $In_xGa_{1-x}$ alloys does not change significantly with the alloy composition. While there is a slight change in the relative energies of the bands (discussed below), the most notable change in the ARPES band structure is a downward shift in the Fermi level with increasing In concentration (**Figure S8**). This can be observed through the shrinkage of the pocket of states located around the K point in both ARPES and the calculated VCA Fermi surface (note that the empirical shift in the VCA Fermi level is aligned to the shape of the larger nearly free electron component of the Fermi surface, not the presence or size of the pocket). The overall shift in the Fermi level is consistent with the reduction in work function as the In concentration increases (Figure S11) and is important for explaining the trend in superconductivity discussed below.

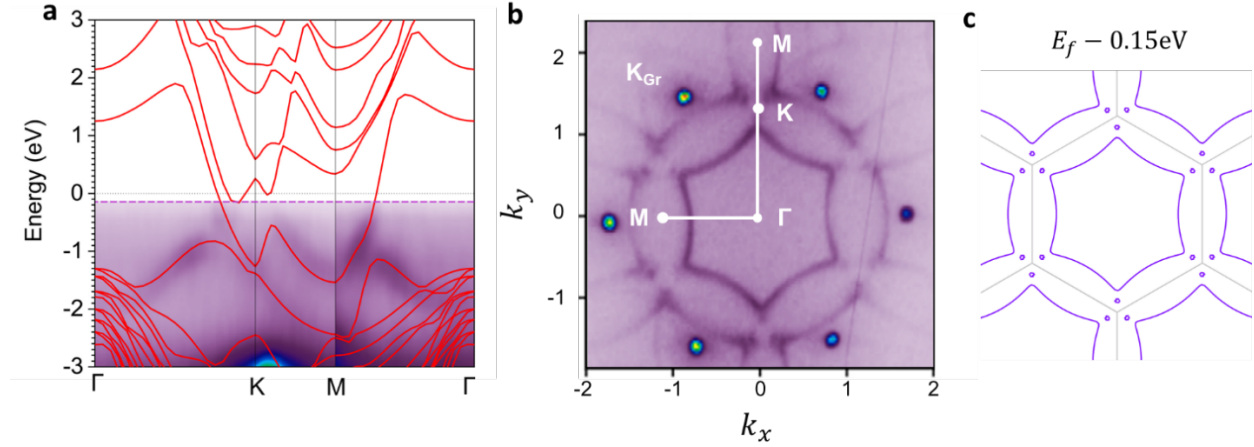

**Figure 4. Electronic structure of 2D-In$_{0.75}$Ga$_{0.25}$ alloy.** (**a**) Calculated band structure of 2L In$_{0.75}$Ga$_{0.25}$ alloy/SiC under VCA (red lines) and measured APRES spectra (purple map) along the high symmetry path Γ–K–M–Γ of the alloy/SiC Brillouin zone. The dashed purple line is the experimental Fermi level, which is 0.15 eV below the calculated Fermi level (likely due to the absence of the graphene cap in the VCA calculation; see Figure S10). (**b**) ARPES-measured Fermi surface where $k_x$ and $k_y$ are the in-plane electron crystal momenta. High symmetry paths and points of the first Brillouin zone of 2L In$_{0.75}$Ga$_{0.25}$ alloy/SiC are labelled by white lines and solid circles, respectively. (**c**) DFT-calculated Fermi surface of 2L In$_{0.75}$Ga$_{0.25}$ alloy/SiC (purple), where the calculated Fermi level ($E_f$) is shifted by −0.15 eV to match the experimental Fermi level. Note how the size and shape of both the large nearly free electron surface and also the small pockets around the K point are simultaneously well-described with a single Fermi level shift. The Brillouin zone of alloy/SiC is shown in grey.

The 2D metal alloy exhibits a composition-dependent dielectric function. **Figures 5a** and **5c** show the dielectric functions determined from variable angle spectroscopic ellipsometry (VASE) in combination with a regression analysis using a multilayer model[13]. We have previously applied this approach to obtain the dielectric functions for 2D Ga and 2D In[13] and obtained good agreement with first-principles calculations. The model consists of the SiC substrate, a modified SiC substrate describing the uppermost SiC layers to which the metal atoms are covalently bonded[13], the 2D metal film, and graphene. The dielectric functions of the 2D metals are well described by a sum of two to three Lorentz functions and a Drude term. The Drude term describes free electron contributions due to intraband excitations ("plasmon mode") and the Lorentz terms describe localized electronic contributions due to interband excitations[28]. In our previous work on 2D Ga and 2D In[13], detailed statistical analysis of the thickness of the 2D metals using cross-sectional transmission electron microscopy indicated that Ga samples are predominantly bilayer while In samples are a mixture of bilayer and trilayer films. The peaks at ~1.7 to 2.0 eV in the measured $\varepsilon_2$ spectra are interband transitions in bilayer films, while the additional peak at ~1.2 eV for In is related to trilayer films. Similar peak positions are observed here for 2D Ga and In. The 2D alloy samples all have a single peak in $\varepsilon_2$ at ~1.7 to 2.0 eV (**Figure 5c**), consistent with our findings that the alloy samples are predominantly bilayers.

Experiment and theory (**Figures 5b** and **5d**) demonstrate that the dominating Lorentz term centered at ~1.7 to 2.0 eV in $\varepsilon_2$ shifts to lower energies with increasing In content, except for the case of pure Ga in the experimental data. The $\varepsilon_2$ spectra computed for the two random configurations for $In_{0.5}Ga_{0.5}$ alloys have peak positions within ±0.15 eV of those obtained using VCA (**Figure S9c, d**). In contrast, a simulation involving phase separated $In_{0.5}Ga_{0.5}$ alloy structures with In (Ga) in the bottom layer and Ga (In) in the top layer gives peak positions of 2.10 and 1.59 eV, which are significantly different from experiment and further support the well-mixed nature of the alloys. The interband transitions at ~2 eV result from quantum-confined bands centered around the K–M high symmetry path in the Brillouin zone[13]. These transitions are not affected by the discrepancies between the computed and experimental Fermi levels in ARPES (see also **Figure S13**).

Previously, the Lorentz resonance was shown to decrease with the effective thickness of the quantum well for bilayer and trilayer Ga and In samples[13]. DFT indicates that the electrostatic potential in 2D In is deeper than that in 2D Ga (**Figure S14**), indicating that the smaller energy separation between the quantum-confined bands in 2D In results primarily from its larger thickness compared to 2D Ga rather than a weaker crystal potential. For the alloys, the interband transition energies decrease as the In composition increases (**Figure S8**), and the dependence of these interband transition energies on the effective quantum well thickness agrees well with that obtained for 2L and 3L In and Ga samples (**Figure S15**). This suggests that the optical absorption peaks in these 2D metal systems can be effectively tuned, not just by the number of layers[13] but also more finely by bottom-up control of the alloy composition.

Tunability is also evident in the real part of the dielectric function, with epsilon near zero (ENZ) regions[13] in the near-IR (NIR) and visible range prominently occurring for the DFT-calculated functions **(Figure 5b)** that are in reasonable agreement with the experimentally determined $\varepsilon_1$. The relatively large deviations from ENZ for the alloys seems to be caused by the strong damping of the Drude term causing significant losses evident by the large positive values in $\varepsilon_2$ in the NIR range that are much less pronounced for 2D-Ga and 2D-In. Possible explanations for the large damping of the free electron contribution described by the Drude term might include alloy scattering, interface roughness, or non-uniform charge transfer from the 2D metals to graphene.

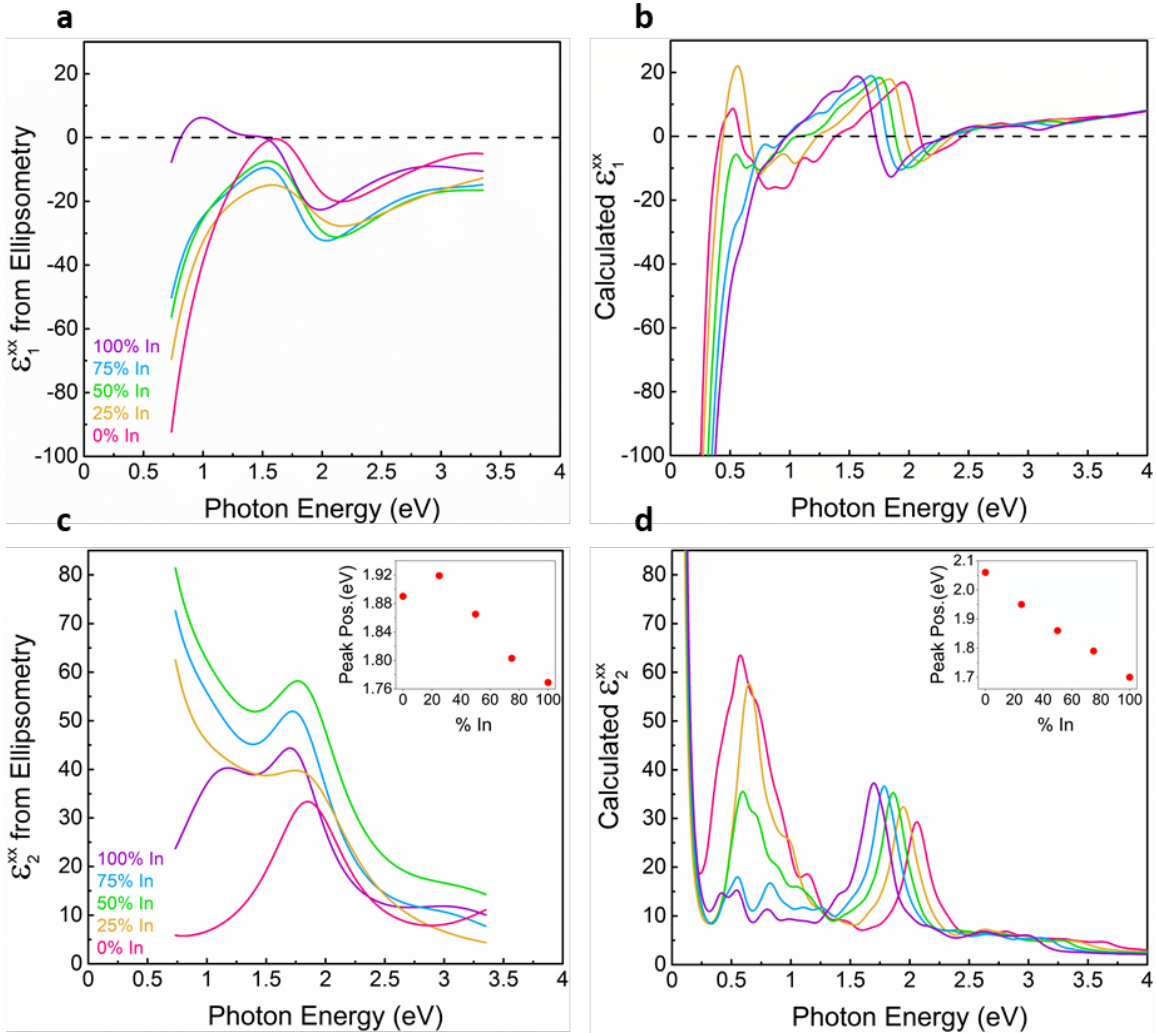

**Figure 5. The complex dielectric function of 2D-In$_x$Ga$_{1-x}$ alloys. (a,b)** Real part of the complex dielectric function $\varepsilon_1^{xx}(E)$, (a) deduced from ellipsometry measurements and (b) computed from first principles (DFT). **(c,d)** Imaginary part of the complex dielectric function $\varepsilon_2^{xx}(E)$, (c) deduced from ellipsometry measurements and (d) computed from DFT. The measurements are obtained using VASE at 55°, 60° and 65°. One model is applied to fit data from all angles simultaneously. We observe an interband transition at roughly 2 eV, which is in excellent agreement to DFT calculations. The insets of panels **(c)** and **(d)** show the major interband optical transition peaks as a function of Indium content. Experimental (computed) peak positions are 1.89 (2.06), 1.92 (1.95), 1.87 (1.86), 1.80 (1.79) and 1.77 (1.70) eV, respectively for 0% In, 25% In, 50% In, 75 % In and 100% In alloys, decreasing with increasing In content for both the VASE measurements and DFT calculations, except for Gallium in the experimental data. A second resonance is evident in the low energy range (~0.7 eV).

The superconducting transition temperature $T_c$ can also be tuned by the alloy composition. **Figure 6a** plots the temperature-dependent resistance of $In_xGa_{1-x}$ alloys, where $x$ ranges from 0 to approximately 0.5. **Figure 6b** plots the extracted $T_c$ as a function of the In concentration $x$. $T_c$ decreases from its pure 2D-Ga value of $T_c$ = 4.1 K[12] to approximately 2 K when the In concentration reaches 50%, with the slope appearing to become steeper around 40% indium. The trend of $T_c$ is consistent with the evolution in Fermi surface geometry in $In_xGa_{1-x}$ alloy. Our prior investigation of the pure-Ga system revealed that contributions to $\alpha^2F$ (the electron-phonon coupling function relevant to superconductivity) are substantial for virtual transitions between the nearly-free-electron-like sheet and the pocket of states around the K point[12]. In **Figure 4** and **Fig. S8**, we combine calculations and ARPES measurements to show that the nearly-free-electron Fermi surface is larger in Ga than it is in In, and the pocket at K disappears as the indium concentration increases. This explains the decrease of $T_c$ with increasing In concentration. The disappearance of the K pockets may be responsible for the change in slope around $x=0.4$.

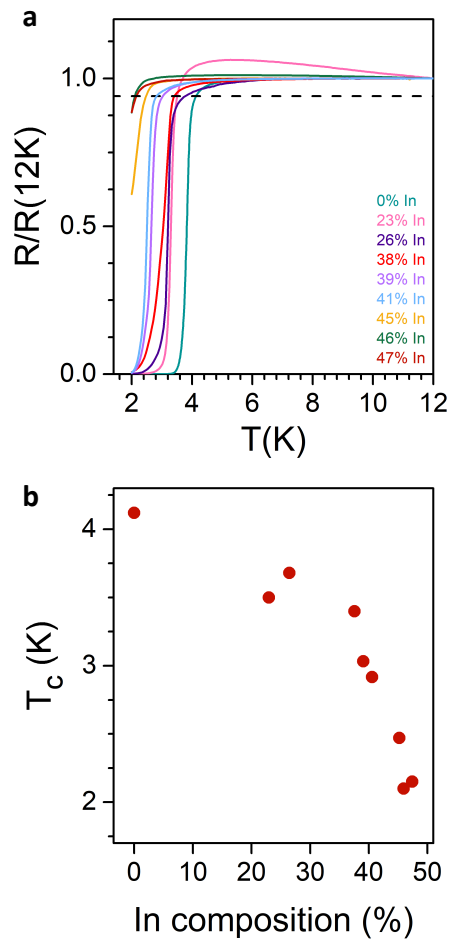

**Figure 6. Superconductivity in CHet-grown 2D-In$_x$Ga$_{1-x}$ alloys. (a)** Normalized resistance vs. temperature $R/R$(12 K) for In concentration $x$ as labeled in the plot. **(b)** The superconducting transition temperature $T_c$ as a function of $x$. We estimate $T_c$ using the black dashed line in (a), where $R(T_c) = 0.94\ R$(12 K). $T_c$ corresponds approximately to the onset of a precipitous resistance drop in the $R(T)$ curves.

**Conclusions**

The formation of environmentally robust, uniform, large scale two-dimensional In$_x$Ga$_{1-x}$ alloys enables detailed chemical and compositional analyses that reveal successful tuning of the alloy composition and associated electronic, optical and superconducting properties, including the degree of charge transfer, the presence and placement of epsilon-near-zero regions, and the Fermi surface topology. The ability to form well-mixed alloys in a 2D geometry that is in intimate contact with a SiC substrate without experiencing substantial compositional layering due to differential bonding strength to SiC likely reflects the *chemical* similarity of Ga and In. Although Ga and In atoms differ in size, this *geometrical* distinction cannot easily be expressed along the layering dimension in an epitaxial bilayer geometry. Similar alloying strategies spanning further across the periodic table may provide opportunities to expand the scope of these environmentally stable 2D metals, with prospects for interlayer or intralayer ordering and associated line phases, enhancements in intercalation kinetics or thermodynamics, and the introduction or enhancement of spin or superconducting order.

**Materials and Methods:**

**Synthesis of 2D-In$_x$Ga$_{1-x}$ alloys.** Atomically thin In$_x$Ga$_{1-x}$ alloys with tunable compositions were achieved via the confinement heteroepitaxy (CHet) in a horizontal tube furnace fitted with a 1-inch outer diameter quartz tube[12]. Various alloy precursors were formed by the mixing of metallic gallium and indium powers with different ratios on a custom-made alumina crucible at 80°C. Annealing at 800°C in 500 Torr Ar environment (Ar flow of 50 sccm) for 30 minutes were determined to be optimum conditions for the CHet process.

**XPS.** XPS measurements were performed with a Physical Electronic Versa Probe II**,** using a monochromic Al K$_\alpha$ X-ray source (hv = 1486.7 eV) and a concentric hemispheric analyzer. High-resolution spectra over a radial area of 200 μm were obtained at a pass energy of 29.35 eV and charged corrected to the graphene peak in C1s spectrum at 284.5 eV. A U2 Tougaard background was used to fit XPS spectra. The $In_xGa_{1-x}$ compositions were calculated based on the In 4d and Ga 3d spectra.

**AES.** AES maps were acquired using a beam current of 10 nA and energy of 10 keV with a Physical Electronics Model 670 Scanning Auger system with a field emitter. A 2-point acquisition method is used for intensity calculation at each point. The following peak/background energy values are used: In: 404.1/420.1 eV, Ga: 1068.3/1081.0 eV, C: 267.8/291.0 eV, Si: 1613.2/1628.0 eV, O: 509.3/532.0. Maps are the average of 5, 20, 5, 15, 15 frames for In, Ga, C, Si and O, respectively.

**Cross-sectional STEM and EELS.** Cross-sections from the $In_xGa_{1-x}$ alloy samples are prepared by a Helios G4 PFIB UXe DualBeam with a $Xe^+$ plasma ion source to avoid contamination from typical gallium FIB. A ~200 nm amorphous carbon coating is deposited by the electron beam at 5 keV and 6.4 nA. Then the $Xe^+$ ion beam is utilized to deposit a 5 μm tungsten layer at 30 KeV. The samples were then prepared by performing a standard lift-out procedure and attaching to TEM half grids using tungsten deposition. Finally, both sides of the samples were thinned in multiple steps by gradually lowering the ion beam voltage from 30 kV to 2 kV until the deposited tungsten is almost consumed, and the cross-section window appeared transparent in the electron beam image at 3 keV. High-resolution STEM imaging and EELS of the prepared cross-sections is performed in an FEI Titan 80–300 HB Cubed transmission electron microscope equipped with a high-brightness gun (XFEG) and two spherical aberration correctors in both the probe and the image forming lenses. Operational conditions are optimized to minimize the electron beam damage at the interface for both imaging and EELS. STEM images are collected at 300 keV and less than 60 e/Å$^2$ /sec electron dose rate using an in-column Fischione HAADF detector (model 3000) at 19.1 beam convergence angle, 50.5–200 mrad collection angles at 115 mm Camera length, and 50 μm C2 aperture. Core-loss EELS maps at the interface are acquired at 200 keV, 91 mm Camera Length, ~ 50 pA screen current using a Gatan K2 IS detector. In order to minimize the drift during the acquisition, the beam was blocked, and the stage was left for 20 minutes to stabilize. Then the EELS map was acquired for 67 seconds only at 0.94 Å pixel size, 0.01 Seconds pixel time, 0.5 eV/channel electron dispersion, and 4 eV FWHM energy resolution. Multi-statistical analysis (MSA) is performed on the EELS spectrum image to denoise the spectrum performs performing mapping and quantification. The quantification of the EELS signal is performed using the Hartree-Slater cross-section method for the whole denoised spectrum image. Then the indium and gallium compositional percentages are normalized to indium to gallium ratio.

**Azimuthal RHEED**. The principle of azimuthal reflection high-energy electron diffraction (ARHEED) has been described elsewhere[29] and has been applied to graphene on $SiO_2$,[29,30] monolayer $MoS_2$ and monolayer $WS_2$ on c-plane sapphire substrates.[31] The glancing incident angle of the high-energy electron beam was oriented at about 1° relative to the sample surface. The glancing electron beam was elongated on the sample surface (at least several mm$^2$) and electrons' short mean free path allowed the near surface structure from a large surface area of a sample be probed. The electron energy and emission current used in $10^{-8}$ Torr vacuum were 20 keV and 45 µA, respectively. In the ARHEED the sample was rotated with a 1.8° step increment around an axis perpendicular to the sample surface. The rotation was controlled by an ultrahigh vacuum compatible step motor. At each azimuthal angle, a RHEED pattern projected on a phosphor screen was recorded by a digital camera placed outside the vacuum chamber. 100 RHEED patterns at 100 azimuthal angles allow the reciprocal space structure above the sample surface to be constructed. The quantitative analysis from either an individual RHEED pattern or the constructed reciprocal space 2D map using Laue conditions gives the in-plane lattice constants and reveals the symmetry of the sample surface, respectively. The scale bar in the RHEED pattern has been calibrated previously by using a CdTe crystal with known lattice constants.[32]

**ARPES**. ARPES measurements were performed at the Microscopic and Electronic STRucture Observatory (MAESTRO) beamline at the Advanced Light Source at Lawrence Berkeley National Lab. The sample was annealed at 550 K for 1 hour in the end-station before measurements to remove adsorbates from the transfer of the sample through air. Measurements of 2D Ga, $In_{0.25}Ga_{0.75}$, $In_{0.50}Ga_{0.50}$, $In_{0.75}Ga_{0.25}$ and In structures were performed using photon energies of 140, 123, 126, 200 and 110 eV, respectively. Photoemission spectra were collected by moving the sample around one angle while using the angle-resolved mode of a Scienta R4000 electron analyzer for the collection of the other angular axis.

**Variable angle spectroscopic ellipsometry (VASE).** VASE measurements are implemented using an M-2000 ellipsometer (*J.A. Woollam*) in rotating-compensator (RCE)-mode. The angle of incidence (AOI) is varied in 5° steps from 55° to 65°. The measurements are conducted at a wavelength range from 300 to 1700 nm (0.74 eV – 3.35 eV). All samples are measured under ambient conditions in a clean room with a macroscopic laser spot size of ~1 mm. Models of the experimental data are obtained using the EP4 modeling program from Accurion GmbH. The samples are all modeled as a layer stack with the following setup: Air – epitaxial graphene (EG) – 2D metal – modified SiC - SiC (substrate). The dispersion functions for air, graphene and SiC are adjusted from literature.[33,34] The 2D metals are modelled by considering the free and bound charge carrier contributions, using a combination of Drude and Lorentz functions[13]. The resulting dielectric function of each sample is extracted by simultaneously fitting the experimental spectra taken for the different AOIs to the model of the layer stack.

**Transport measurements.** Four-probe resistance measurements were performed in a Quantum Design physical property measurement system (PPMS) using an excitation current of 1 μA. Electrical contacts are made by soldering Indium dots onto the EG/In$_x$Ga$_{1-x}$/SiC heterostructures in a colinear four-point-probe configuration parallel to the steps. Contact pitches were on the order of hundreds of micrometers.

**Theory**

**Phase stability calculations**. All DFT calculations on phase stabilities were performed using the GGA-PBE[35] exchange-correlation functional and the PAW pseudopotentials. Plane-wave expansions were truncated at an energy cutoff of 500 eV. All structural relaxations were performed using dipole corrections to the total energy and to the electrostatic potential in the out-of-plane direction, until the remaining forces were within 0.001 eV Å$^{-1}$. Brillouin zone samplings were performed on grids with k-point densities equivalent to that of a 30×30×1 grid for a 1×1 Ga/SiC unit cell. Graphene/Ga/SiC calculations were performed using five repeating units along the *z* direction for a 2×2 bilayer graphene plus a √3×√3 R30º Ga/SiC supercell, capped from below with H, and also for a 2×2 supercell of the same used for most of the alloy calculations to achieve correct dopant density. The Ga$_{Si}$Ga$_C$ stacking (with Ga replaced by In in the appropriate alloys) was used for all calculations.

**Band structures and dielectric functions.** DFT calculations within the virtual crystal approximation (VCA)[27] were performed using the local density approximation (LDA)[36] for the exchange-correlation functional as implemented in the plane-wave pseudopotential code, QUANTUM ESPRESSO[37]. We show in **Figure S9c** that the band structures obtained with LDA and PBE exchange-correlation functionals are nearly identical; structural parameters are likewise similar, with the interlayer distances and the SiC lattice constant being slightly larger for PBE than for LDA (**Table S3**). Optimized norm-conserving pseudopotentials[38] were used to generate the pseudopotentials of virtual atoms that describe on average the electron-ion interactions in alloys. Three concentrations of Ga, 25%, 50% and 75% Ga, were considered to compare with experimental samples. A plane-wave kinetic energy cut-off of 90 Ry was used, and an energy threshold of $10^{-10}$ Ry was applied for the self-consistent cycle. A Monkhorst–Pack k-point mesh of 26×26×1 was used for geometry optimization and self-consistent field calculations, while a denser k-point mesh of 52×52×1 was needed to obtain density of states and converged dielectric functions. SiC substrates are modeled by six SiC layers passivated by hydrogen atoms where the bottom SiC layer is fixed during the structural relaxation. All other atomic positions are relaxed until the forces acting on each atom are less than 0.0001 Ry/Bohr. To eliminate interactions between slabs along the z direction due to the periodic boundary condition, a vacuum length of 16 Å with an electrostatic dipole correction is employed.

The complex dielectric function was computed using the independent particle approximation[39], where excitonic effects are neglected. This approximation is reasonable given the large in-plane screening present in the 2D metal systems. The volume of the "optically active region" used to normalize the computed dielectric function consists of the metal layers and top three SiC layers. In the sum over states, we included 70 bands involving all valence states to converge $\varepsilon^{\alpha\beta}$. Additionally, the intraband and interband broadening parameters were taken to be, respectively, 0.01 eV and 0.08 eV. These broadening parameters have a negligible influence on the optical peak positions[13]. Band unfolding for random alloy systems was performed using the BandUP code[40,41].

**Acknowledgments**

Funding for this work was supported by Air Force Office of Scientific Research (AFOSR) grant FA9550-19-1-0295, the National Science Foundation (NSF) under DMR-2002651 and DMR-2011839 through the Penn State MRSEC Center for Nanoscale Science, the 2D Crystal Consortium NSF Materials Innovation Platform under cooperative agreement DMR-1539916, the Singapore National Research Foundation, Prime Minister's Office, under its medium-sized center program, the Deutsche Forschungsgemeinschaft (DFG, German Research Foundation) under Germany's Excellence Strategy-EXC2089/1-390776260, the Empire State Development's

Division of Science, Technology and Innovation (NYSTAR) through Focus Center-New York contract C150117, and Horiba corporation. This research used resources of the Advanced Light Source, which is a DOE Office of Science User Facility under contract no. DE-AC02-05CH11231. The electron microscopy work was funded by the US AFOSR Award FA9550-19-1-0239 and the NSERC - Natural Sciences and Engineering Research Council of Canada Discovery Grant program. The STEM/EELS work is performed at the Canadian Centre for Electron Microscopy (CCEM), McMaster University. Optical properties computations were performed on the NUS Graphene Research Centre cluster and National Supercomputing Centre Singapore (NSCC). We acknowledge Jeffery Shallenberger for help with XPS analysis and Vince Bojan for AES support.

**Author Contributions**

S.R. and J.A.R. conceived the idea and directed the research. S.R. carried out the synthesis of 2D-In$_x$Ga$_{1-x}$ alloys and conducted XPS, AES and SEM. STEM and EELS characterization is performed by H.E-S. under the supervision of N.B. B.N.K performed phase stability calculations and explicit alloy band structure calculations under the direction of V.H.C. X.C. performed and analyzed ARHEED under the supervision of G.-C.W. K.N. and M.L. performed VASE measurements and modelling of the ellipsometric spectra under the supervision of U.W. R.K., A.B., and C.J. performed ARPES measurements under the direction of E.R. S.R analyzed the ARPES data with significant input from R.K. W.H. performed the calculations on optical properties and band structures, and contributed to analyzing the ARPES data, under the supervision of S.Y.Q. C.L performed the transport measurements under the supervision of J.Z. All authors contributed to analyzing and discussing the data. S.K., A.V. and J.A.R. wrote the paper with significant input from W.H., B.N.K, C.L., J.Z., G.-C.W., N.B., U.W., S.Y.Q., V.H.C., and A.C.T.vD.

# Supplementary Information

# Tunable Two-Dimensional Group-III Metal Alloys

Siavash Rajabpour[1,2], Alexander Vera[2,3], Wen He[4,5], Benjamin N. Katz[2,6], Roland J. Koch[7], Margaux Lassaunière[8,9], Xuegang Chen[10], Cequn Li[2,6], Katharina Nisi[8,11], Hesham El-Sherif[12], Maxwell T. Wetherington[2,3,13], Chengye Dong[2,14], Aaron Bostwick[7], Chris Jozwiak[7], Adri C.T. van Duin,[1,2,3,13,14,15,16,17], Nabil Bassim[12,18], Jun Zhu[2,6], Gwo-Ching Wang[10], Ursula Wurstbauer[8,9], Eli Rotenberg[7], Vincent Crespi[2,3,6,13,14,16], Su Ying Quek[4,5,19,20], Joshua A. Robinson.[2,3,13,14]*

21. Department of Chemical Engineering, The Pennsylvania State University, University Park, PA, USA
22. Center for 2-Dimensional and Layered Materials, The Pennsylvania State University, University Park, PA, USA
23. Department of Materials Science and Engineering, The Pennsylvania State University, University Park, PA, USA
24. Department of Materials Science and Engineering, National University of Singapore, 9 Engineering Drive, Singapore, Singapore
25. Centre for Advanced 2D Materials, National University of Singapore, 6 Science Drive 2, Singapore, Singapore
26. Department of Physics, The Pennsylvania State University, University Park, PA, USA
27. Advanced Light Source, Lawrence Berkeley National Laboratory, Berkeley, California, USA
28. Institute of Physics, University of Münster, Münster, Germany
29. Center for Soft Nanoscience, University of Münster, Münster, Germany
30. Department of Physics, Applied Physics and Astronomy, Rensselaer Polytechnic Institute, Troy, NY, USA
31. Physics Department, Technical University of Munich, Garching, Germany
32. Department of Materials Science and Engineering, McMaster University, Hamilton, Ontario, Canada
33. Materials Research Institute, The Pennsylvania State University, University Park, PA, USA
34. 2-Dimensional Crystal Consortium, The Pennsylvania State University, University Park, PA, USA.
35. Department of Mechanical Engineering, The Pennsylvania State University, University Park, PA, USA
36. Department of Chemistry, The Pennsylvania State University, University Park, PA, USA.
37. Department of Engineering Science and Mechanics, The Pennsylvania State University, University Park, PA, USA
38. Canadian Centre for Electron Microscopy, Hamilton, Ontario, Canada
39. Department of Physics, National University of Singapore, Singapore, Singapore
40. NUS Graduate School Integrative Sciences and Engineering Programme, National University of Singapore, Singapore 117456

* jrobinson@psu.edu

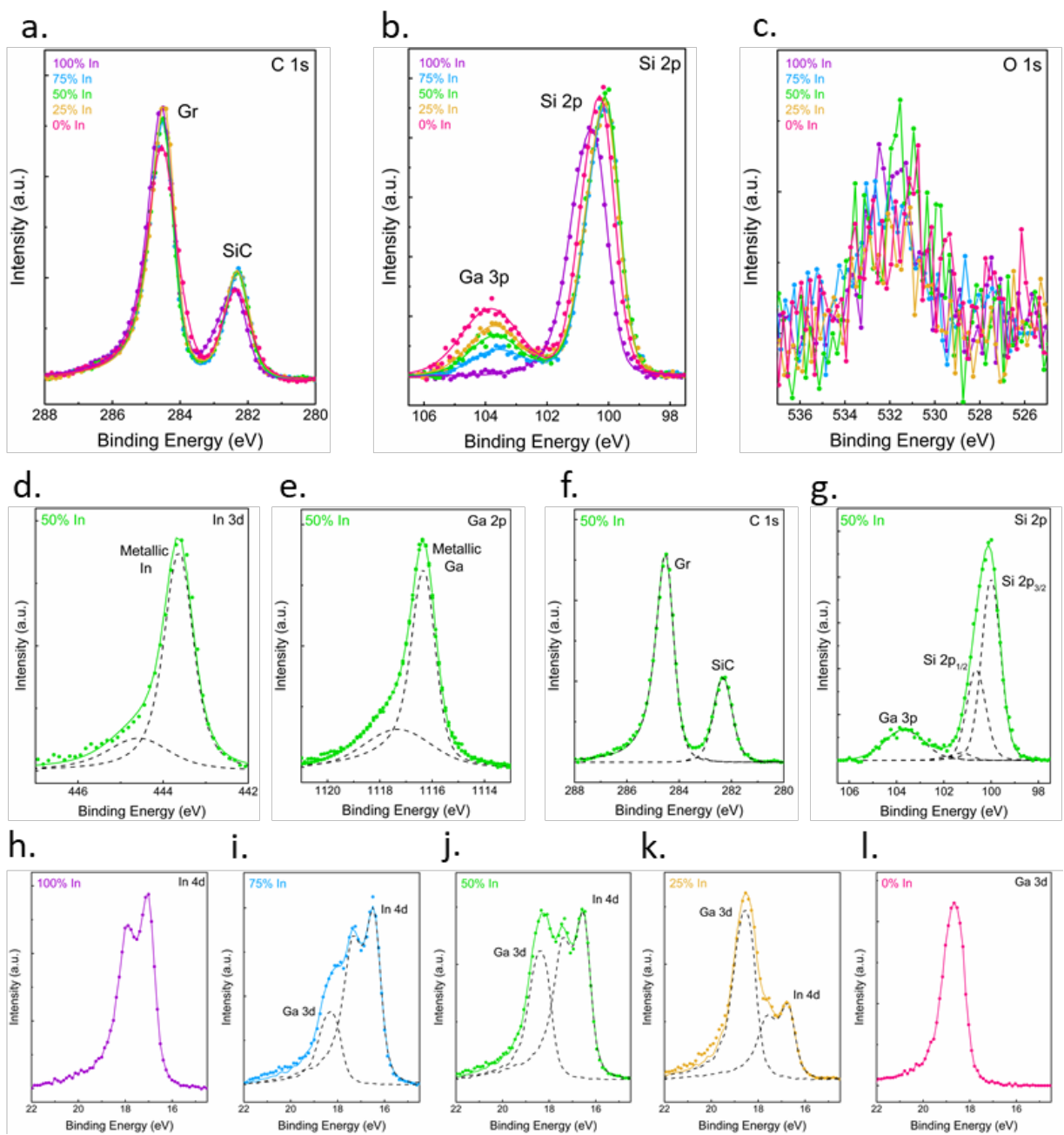

**Figure S3. X-ray photoelectron spectroscopy (XPS) captured for the 2D-In$_x$Ga$_{1-x}$ alloys.** **(a)** The split in C 1s for various 2D alloys verifies the existence of the intercalant at the interface of SiC/EG **(b)** Slight shift in Si 2p peak positions are correspond to different neighbor's chemistries. **(c)** The origin of O 1s signals is likely due to regions of oxidized surface Ga and In particles that are not intercalated. **(d-g)** show the 2D-In$_{0.5}$Ga$_{0.5}$ alloy as a representative sample. The higher binding energy peak in the Ga 2p and In 3d regions correspond to gallium and indium oxide, indicating that some Ga and In remaining on the surface of the EG and subsequently oxidize following exposure to ambient. **(h-l)** For higher accuracy, Ga 3d and In 4d is used to calculate the composition of the 2D-In$_x$Ga$_{1-x}$ alloys. Ga 3d spectrum in 2D-Ga **(l)** and In 4d spectrum in 2D-In **(h)** are utilized to deconvolute the Ga 3d and In 4d regions for alloys. The spectra are charge referenced to the sp$^2$ C (Gr) peak at 284.5 eV.

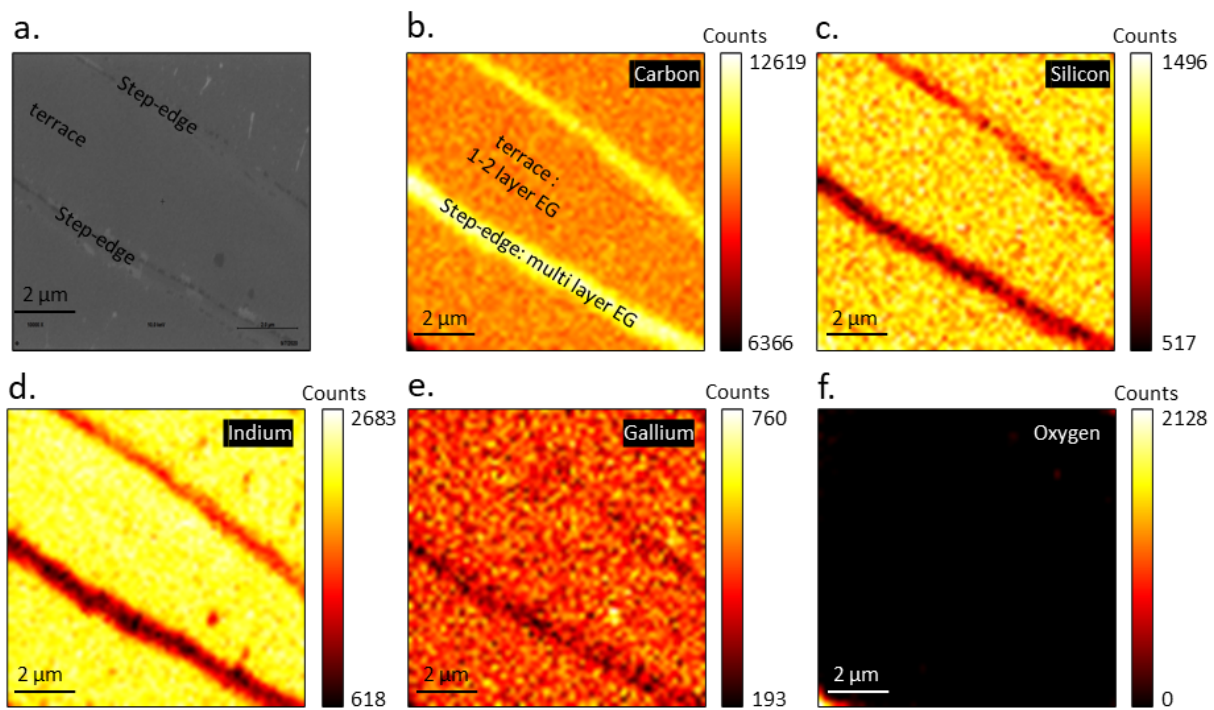

**Figure S4. Scanning electron microscope image (a) and corresponding Auger electron spectroscopy (AES) maps (b-f) of a 2D-In$_{0.5}$Ga$_{0.5}$ alloy.** AES maps indicate (e) Ga and (d) In atoms are uniformly dispersed across a 10 μm × 10 μm region without lateral segregation. The diagonal stripes in which the C signal (b) is stronger is attributed to the greater number of EG layers at the step edges compared to the terrace regions. The greater number of EG layers attenuates the Ga, In and Si signals as these elements sit beneath the EG. The O map indicates that the oxygen content in the alloy is minimal, suggesting the EG protects the alloy from oxidation.

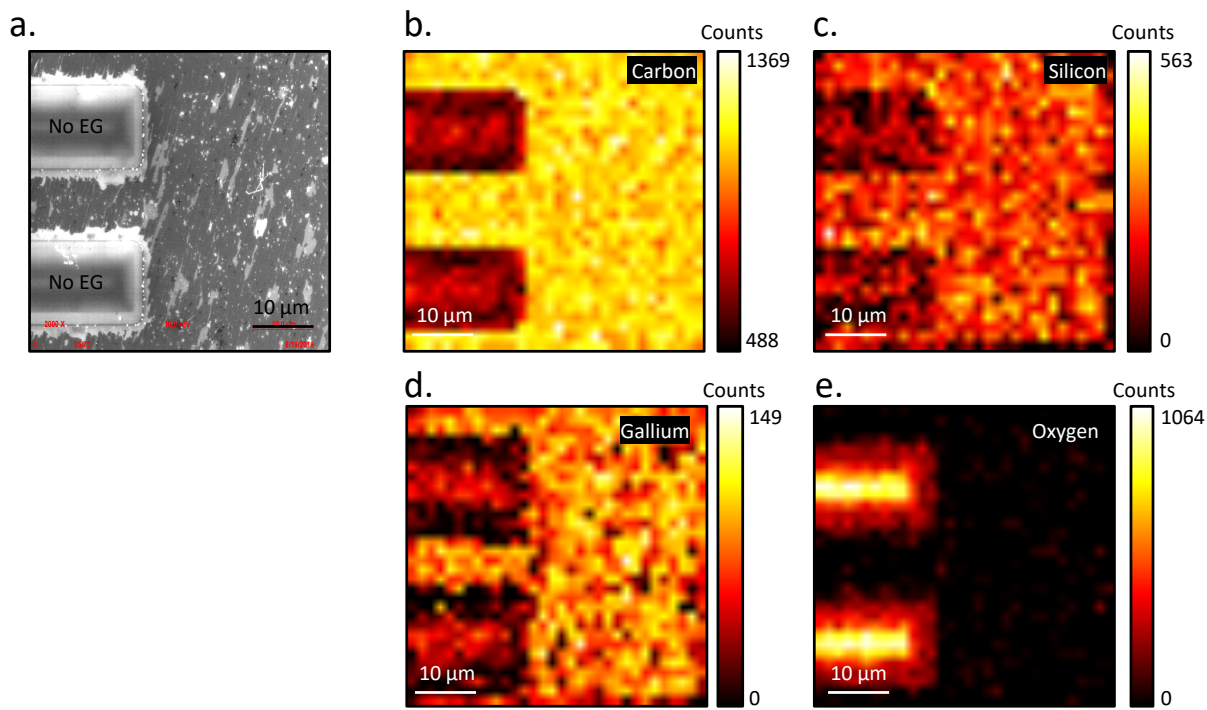

**Figure S3. Oxidation of a patterned 2D-Ga sample.** SEM (a) demonstrates the location where the graphene is selectively removed by plasma etching ("No EG" region), and AES maps (b-e) verifies the critical role of EG as the protective capping layer for CHet-derived 2D metals. In the regions without EG, there is a significant amount of oxidation, while the presence of EG protects the 2D metals form the oxidation and the oxygen content in the metal is below the detection limit.

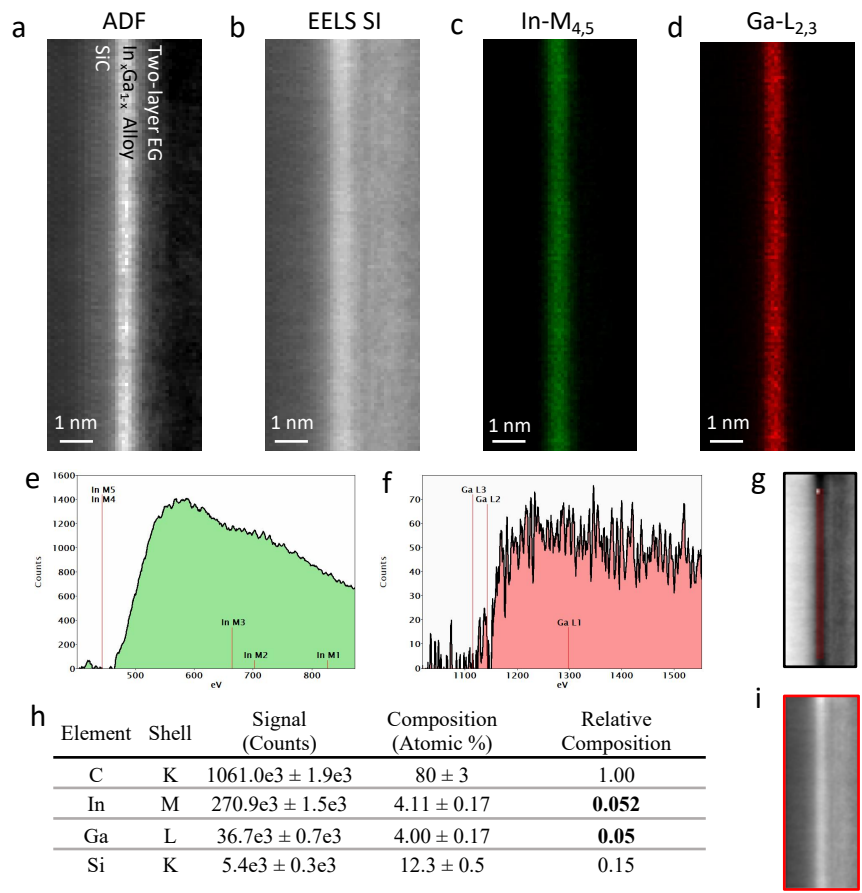

**Figure S4. EELS quantification of the 2D-In$_x$Ga$_{1-x}$ alloy.** (a) Analog Annular dark-field (ADF) image of bilayer In$_x$Ga$_{1-x}$ at the SiC/EG interface. (b) EELS spectrum image (EELS SI) acquired simultaneously with the ADF. Both (a) and (b) have 48 pixels ×130 pixels with ~1 Å pixel size. To achieve spatially resolved signals within the atomic scale. EELS signal is collected by Gatan's K2 IS direct electron detector. (c) and (d) are the Indium and Gallium EELS maps of the In-M$_{4,5}$ and Ga-L$_{2,3}$ edges. The EELS mapping is conducted using Digital-Micrograph software after refining the EELS spectrum image using principal component analysis (PCA). (e) and (f) are the In-M$_{4,5}$ and Ga-L$_{2,3}$ edges after background subtraction. The edge signals are extracted from the bi-layer In$_x$Ga$_{1-x}$ alloy region in (g). The vertical axis represents the electron counts as measured by the direct electron detector. The Ga-L$_{2,3}$ signal is noisier than the In-M$_{4,5}$ because its location is at higher energies (1115 eV) where the total signal is weak, while the In is a delayed maximum at 443 eV. (h) Table summary of the quantified EELS spectrum image performed using the Hartree-Slater cross-section method performed using Digital-Micrograph software. The quantification is applied to all the present elements in the whole spectrum image volume as shown in (i). The relative compositions of Indium and Gallium show a 1:1 ratio, which implies that the In$_x$Ga$_{1-x}$ alloy is In$_{0.5}$Ga$_{0.5}$

**Azimuthal RHEED.** Before the measurement of the 2D map of EG/2D-In$_{0.5}$Ga$_{0.5}$/SiC, the 2D maps of a SiC substrate and bilayer graphene on H terminated SiC substrate were measured using azimuthal reflection high-energy electron diffraction (ARHEED) at room temperature in a high vacuum chamber without any sample treatment. Figures S4(a) and (b) show the 2D maps of pure SiC and bilayer graphene on SiC, respectively. A 2D map is constructed and expressed as functions of momentum transfer parallel to a sample surface $\mathbf{k}_\parallel$ over 360° in-plane azimuthal angles φ at a constant momentum transfer perpendicular to a sample surface $\mathbf{k}_\perp$.[1] The momentum transfer $\mathbf{k}$ = $\mathbf{k}_{out}$ - $\mathbf{k}_{in}$, where $\mathbf{k}_{out}$ and $\mathbf{k}_{in}$ are electron's outgoing and incoming wave vectors, respectively. $|\mathbf{k}_\parallel|$ = $\sqrt{k_x^2 + k_y^2}$, where $k_x = \mathbf{k}_\parallel \sin φ$ and $k_y = \mathbf{k}_\parallel \cos φ$. Each concentric circle represents $|\Delta \mathbf{k}_\parallel|$ = 2 Å$^{-1}$ radially outward. The blue and black grids are constructed from the measured reciprocal space unit vectors $\mathbf{a}_s^*$ and $\mathbf{b}_s^*$ for SiC and measured $\mathbf{a}^*$ and $\mathbf{b}^*$ for bilayer graphene on SiC, respectively. Both show a six-fold symmetry. The relative rotation of unit vectors between graphene and SiC is 30°. The real space in-plane lattice constant $a$ for a hexagonal lattice van be obtained from measured reciprocal space vector $|G(hk)| = \frac{4\pi}{\sqrt{3}a}\sqrt{h^2 + hk + k^2}$, where $a$ is a real space lattice constant.

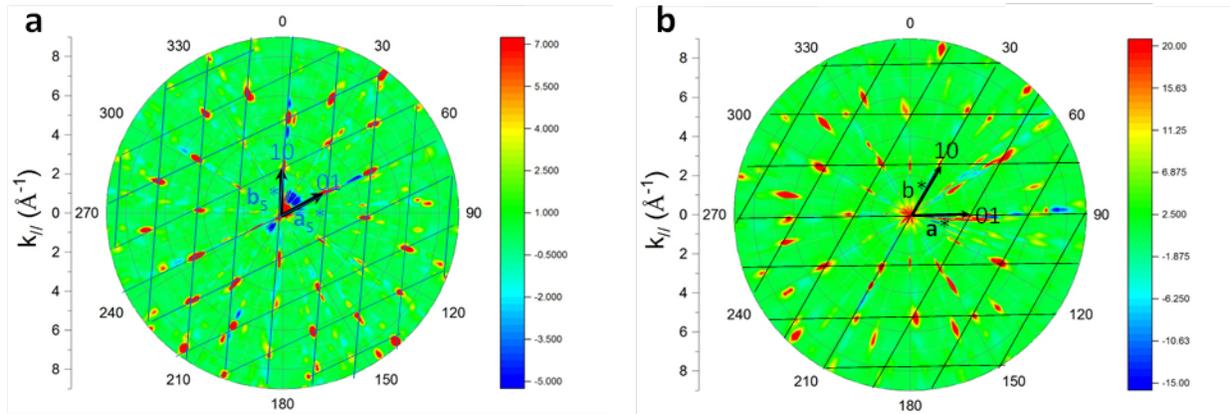

**Figure S5. 2D maps constructed from the measured RHEED patterns at 100 different azimuth angles at room temperature from SiC and bilayer graphene on SiC.** (a) The measured lengths of reciprocal space unit vectors $\mathbf{a}_s^*$ (= $G(01)$) and $\mathbf{b}_s^*$ (= $G(10)$) from SiC are 2.36 ± 0.03 Å$^{-1}$. This measured 2.36 ± 0.03 Å$^{-1}$ can be converted to a real space in-plane lattice constant $a_s$ = 3.08 ± 0.04 Å. This measured reciprocal space value is close to the theoretical SiC's reciprocal space unit vector length of 2.361 Å$^{-1}$ using a real space bulk lattice constant of 3.08 Å. The intersections of blue grid lines cover almost all diffraction spots of SiC. Weak spots not located at the intersections of grid lines are from Kikuchi spots or lines. The $\mathbf{k}_\perp$ of this 2D map is at 6.64 Å$^{-1}$. (b) The measured lengths of reciprocal space unit vectors $\mathbf{a}^*$ and $\mathbf{b}^*$ from the bilayer graphene are 3.03 ± 0.04 Å$^{-1}$. This length is 25% larger than that from SiC. This measured 3.03 ± 0.04 Å$^{-1}$ can be converted to a real space in-plane lattice constant $a$ = 2.39 ± 0.03 Å. The measured reciprocal space value is close to theoretical graphene unit vector length of 2.948 Å$^{-1}$ using a real space bulk graphite lattice constant of 2.46 Å. The intersections of black grid lines cover diffraction spots from graphene. Note that additional spots can be accounted for the blue grids from SiC in (a). The left-over weak spots do not locate at the intersections of grid lines are from Kikuchi spots or lines. The $\mathbf{k}_\perp$ of this 2D map is at 5.63 Å$^{-1}$.

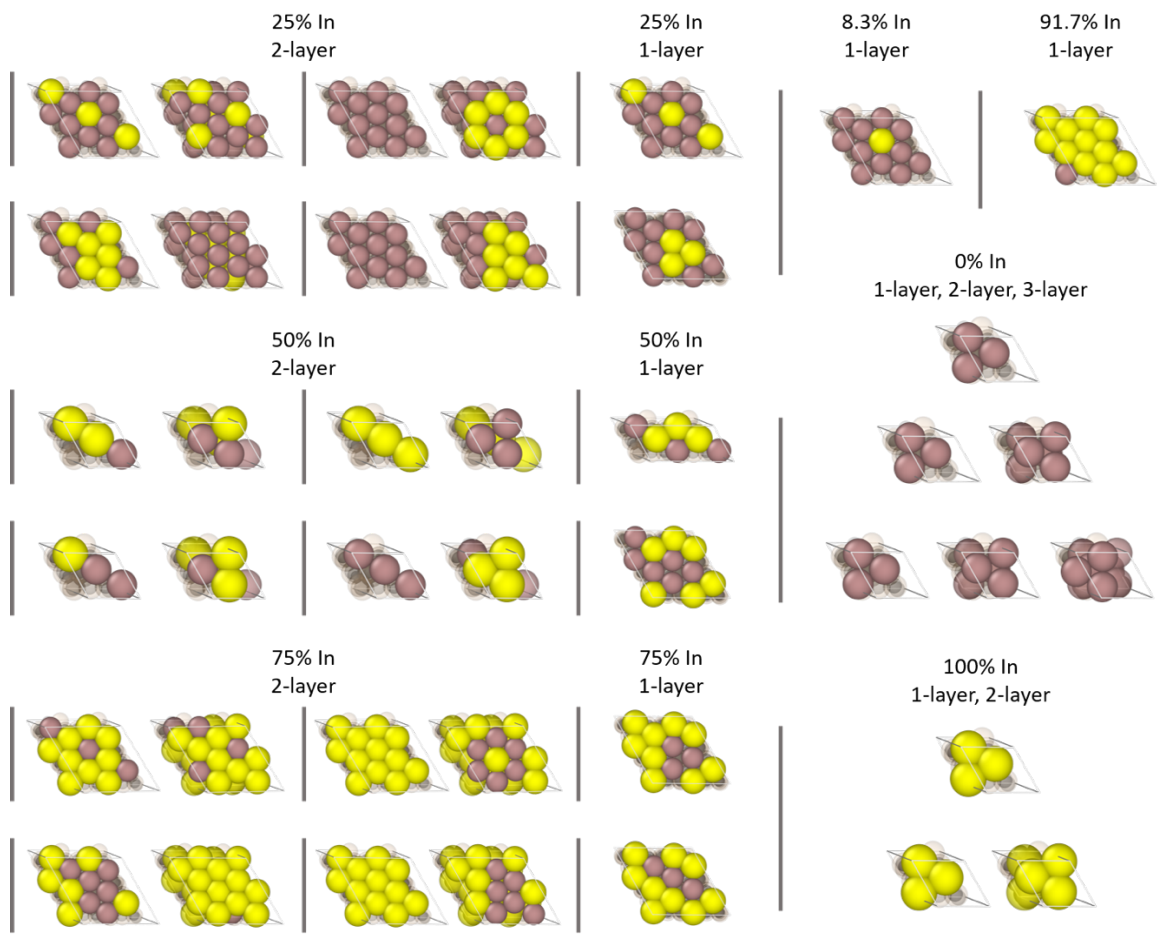

**Figure S6.** Atomic structures of the explicit alloys examined for the computational phase diagram and simple cluster expansion. For multilayer alloys, structures are represented by multiple images, stripping away upper layers to show the structures of those underneath, from right to left. Gallium is reddish brown; indium is yellow. All calculations also incorporated a cap of bilayer graphene, which has been removed for visual clarity. The geometry of the SiC slab underneath is similar to that of **Figure S7** with one less layer of SiC.

**Table S1. Energy error in model fit to DFT energies of structures.** The energy of all twenty-five structures given in **Figure S6** modeled in DFT was modeled by a nearest-neighbor cluster expansion, and the errors in that fit are given here. Of note is that while the model fails badly if no distinction between first and higher level atoms is given, it appears to be successful without distinguishing between 2$^{nd}$- and 3$^{rd}$-layer atoms. This fitting does not include the alloys computed with VCA.

| Alloy fraction of In | Error in model fit (eV) | Alloy fraction of In | Error in model fit (eV) |
|---|---|---|---|
| 0/1, 1L | -0.0243 | 0/1, 2L | 0.0025 |
| 1/12, 1L | -0.0124 | 1/4, 2L | -0.0086 |
| 1/4, 1L | -0.0003 | | -0.0021 |
| | 0.0073 | | 0.0103 |
| 1/2, 1L | 0.0050 | | 0.0000 |
| | 0.0124 | 1/2, 2L | -0.0049 |
| 3/4, 1L | 0.0043 | | -0.0080 |
| | 0.0039 | | -0.0119 |
| 11/12, 1L | -0.0150 | | 0.0119 |
| 1/1, 1L | 0.0000 | 3/4, 2L | -0.0069 |
| 100%-2L | -0.0153 | | -0.0028 |
| 0/1, 3L | 0.0117 | | 0.0117 |
| | | | 0.0000 |

**Table S2. Fitting parameters for the nearest-neighbor cluster expansion.** The parameters for the nearest-neighbor cluster expansion fit to the DFT energies.

| Parameter | Fitted value (eV) |
|---|---|
| Ga_self | -13.1247 |
| In_self | -2.9814 |
| 1$^{st}$ layer Ga-1$^{st}$ layer Ga | -28.9144 |
| 1$^{st}$ layer In-1$^{st}$ layer In | -32.2274 |
| 1$^{st}$ layer Ga-1$^{st}$ layer In | -30.5743 |
| Ga-Si | -15.2819 |
| In-Si | -15.2901 |
| Ga-Graphene | -15.2746 |
| In-Graphene | -15.2974 |
| Upper layer Ga-Upper layer Ga | 1.6800 |
| Upper layer In-Upper layer In | -1.6489 |
| Upper layer Ga-Upper layer In | -0.0089 |
| 1$^{st}$ layer Ga-Upper layer Ga | 1.4772 |
| 1$^{st}$ layer In-Upper layer In | 1.6951 |
| 1$^{st}$ layer Ga-Upper layer In | 1.5701 |
| 1$^{st}$ layer In-Upper layer Ga | 1.5836 |

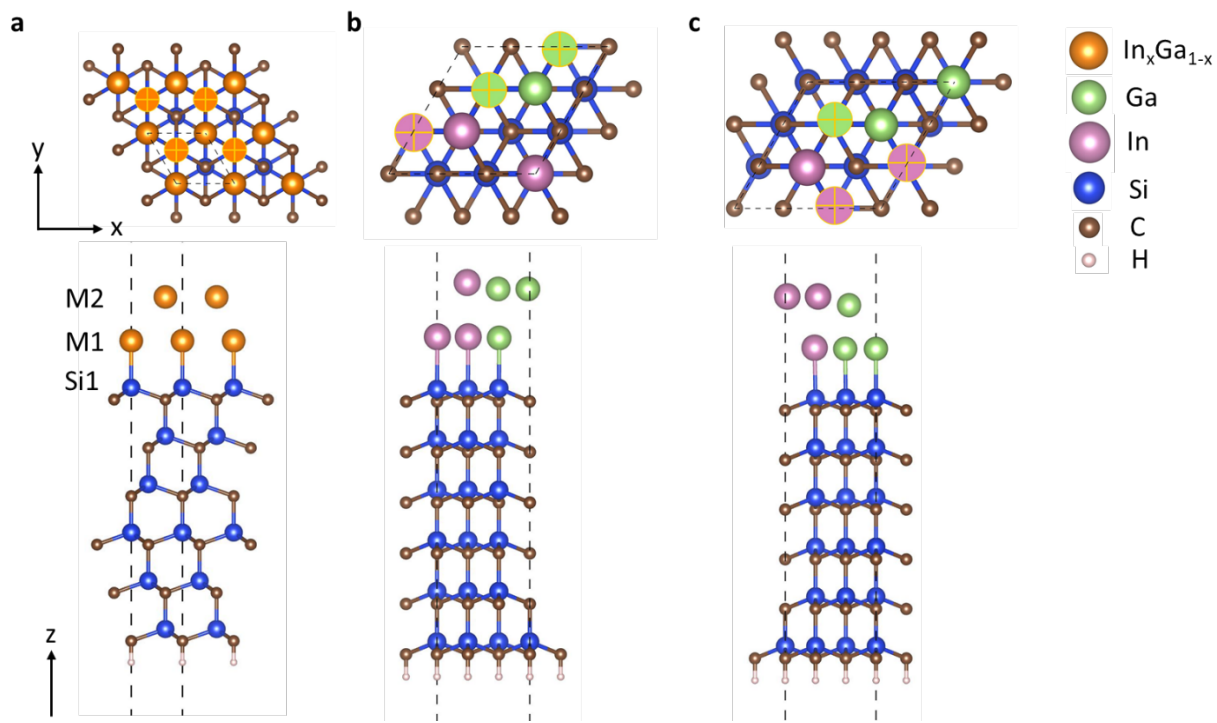

**Figure S7. Atomic structures of 2D In$_x$Ga$_{1-x}$/SiC slabs. (a)** Virtual crystal approximation (VCA) model. Metal atoms are modeled as a mixture of In and Ga. **(b-c):** Models of In$_{0.5}$Ga$_{0.5}$/SiC alloys without the VCA (In and Ga atoms are denoted by pink and green balls, respectively), with an in-plane supercell of r3 × r3. (The models chosen here are the low-energy structures for the r3 × r3 supercells; the structure in (c) is more stable than that in (b) by ~2 meV/atom. See **Figure S6** for other structural models.) Dashed lines indicate the cell boundaries. In all structures, atoms of the top metal layer are marked with a yellow cross and the bottom of the SiC slabs are passivated by hydrogen atoms. The band structures of the models in (b) and (c) are similar to each other and to that for the VCA model (see **Figure S9**).

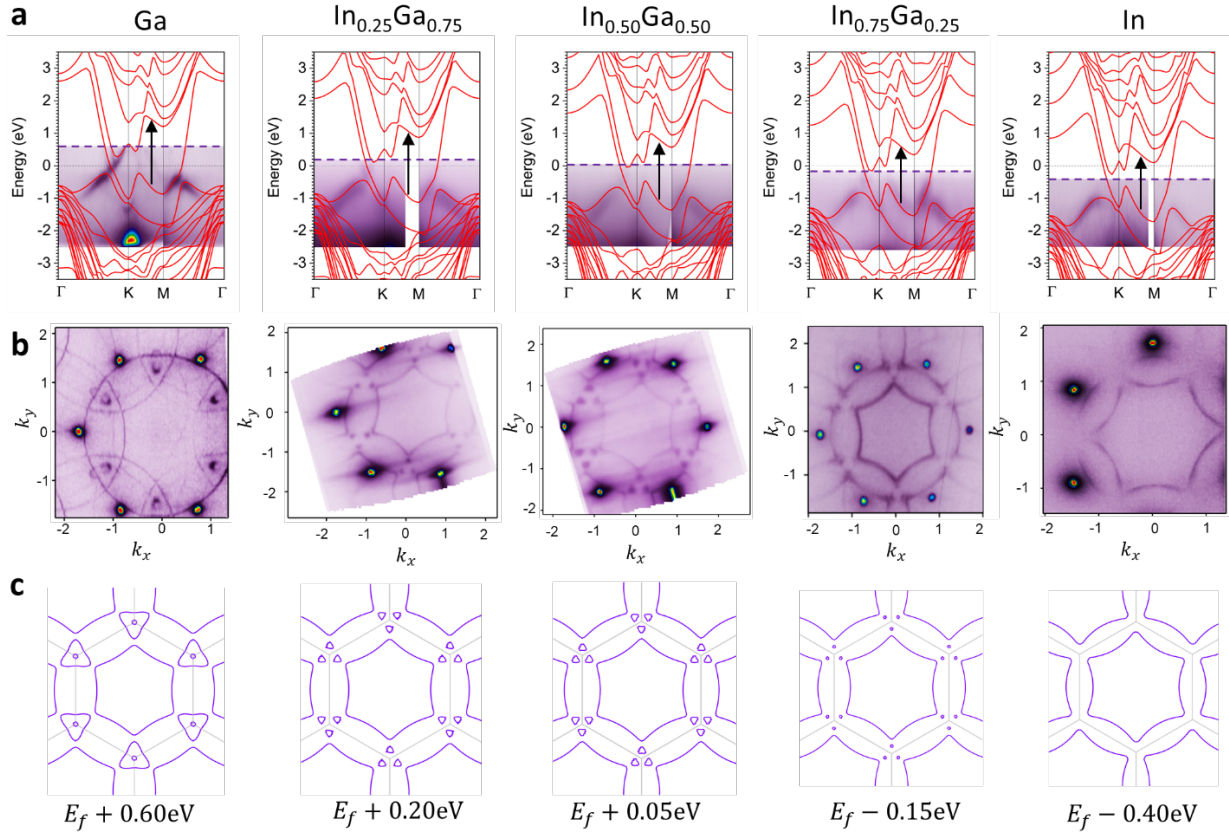

**Figure S8. Electronic properties of 2L In$_x$Ga$_{1-x}$/SiC (x = 0, 0.25, 0.50, 0.75 and 1).** (**a**) Calculated band structures under VCA (red lines) and ARPES measured band structures (purple maps) of 2L In$_x$Ga$_{1-x}$/SiC. The dashed purple line in each band structure is the experimental Fermi level that is deduced using the ARPES-measured Fermi surfaces and band structures. Black arrows mark the interband transitions along the K-M path which contribute to the peak in $\varepsilon_2$ at ~ 1.5-2.0 eV. The corresponding interband transition energies decrease as the In concentration increases. (**b**) ARPES-measured Fermi surface where $k_x$ and $k_y$ are the electron crystal momenta in the in-plane directions. (**c**) DFT-calculated Fermi surface for 2L In$_x$Ga$_{1-x}$/SiC (purple) for the Fermi level deduced from experiment. The BZ is plotted in grey. $E_f$ is the calculated Fermi level.

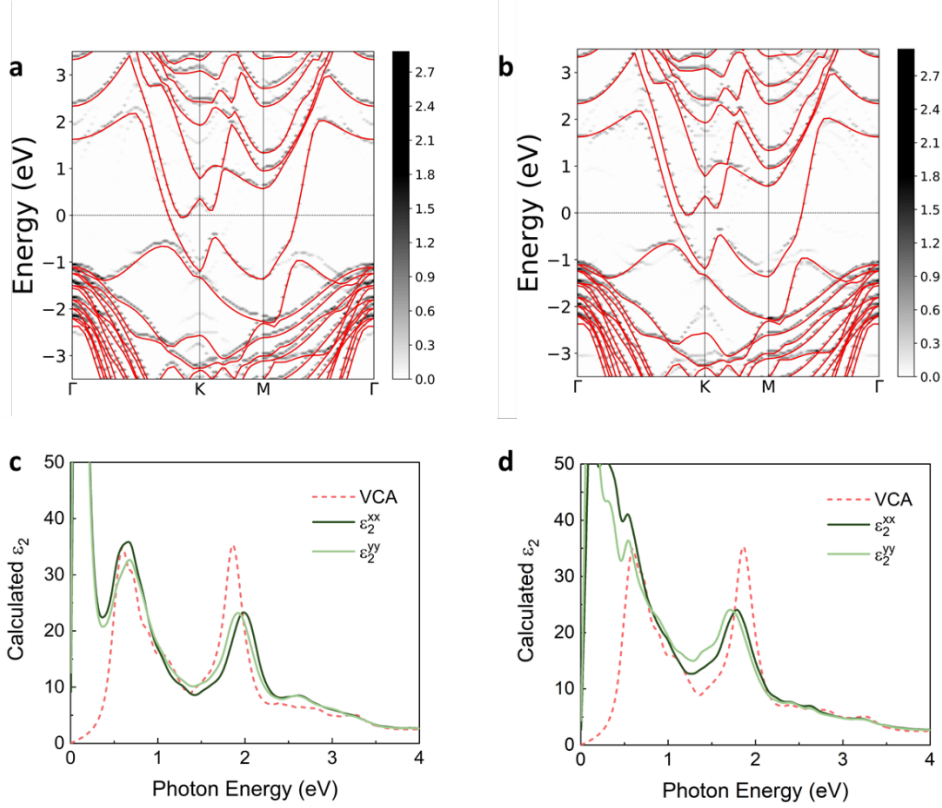

**Figure S9. Electronic and optical properties of 2D In$_{0.5}$Ga$_{0.5}$ with different structural models. (a-b)** Band structure for VCA model (red lines) plotted with effective primitive cell band structures (gray projections) for two instantiations of In$_{0.5}$Ga$_{0.5}$ alloys (as explicit alloys without using the VCA) in (a) Figure S7b and (b) Figure S7c. The gray scale represents the spectral function resulting from the band-unfolding procedure[2]. The Fermi level is at 0eV. **(c-d)** Interband contributions to the imaginary parts of the dielectric functions for the In$_{0.5}$Ga$_{0.5}$ alloys (no VCA) in (c) Figure S7b and (d) Figure S7c (solid lines). Dashed lines: corresponding result from the VCA model ($\varepsilon_2^{xx} = \varepsilon_2^{yy}$). The peak position close to 2 eV does not change significantly across the models (within ±0.15 eV compared to that computed by VCA).

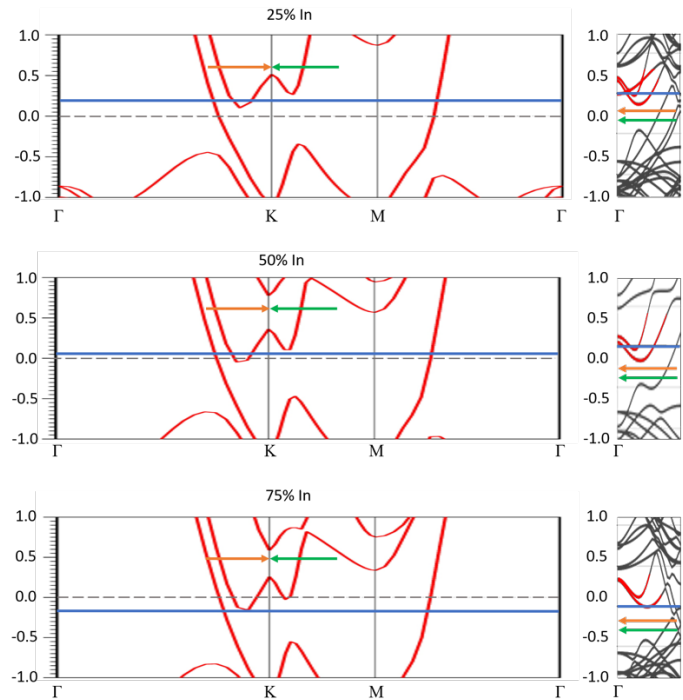

**Figure S10. Comparison of Fermi levels in uncapped VCA alloys (with and without the empirical shift) and atomistic alloys with a graphene bilayer cap.** The left-hand column depicts the VCA bandstructures with the original computed Fermi level as a dashed line and the Fermi level empirically shifted to match the ARPES bandstructure in blue. To minimize computational cost, in these calculations there is no graphene cap. The right-hand column depicts explicit alloy calculations within a supercell with a graphene bilayer cap. The horizontal axes have been scaled to match the scale of the VCA calculation, and the colored arrows show the corresponding k-space directions for the two pockets in the folded (explicit) and unfolded (VCA) bandstructures. To facilitate comparison, the relative alignment of the two sets of bandstructures has been set to align the bottoms of the upper Fermi pockets (i.e. those tracked by the green arrow). The inclusion of the graphene bilayer in the explicit alloy calculation brings the Fermi level reasonably close to that set by empirical shift to match ARPES, providing evidence that charge transfer between metal and graphene is the origin of the shift in the Fermi level.

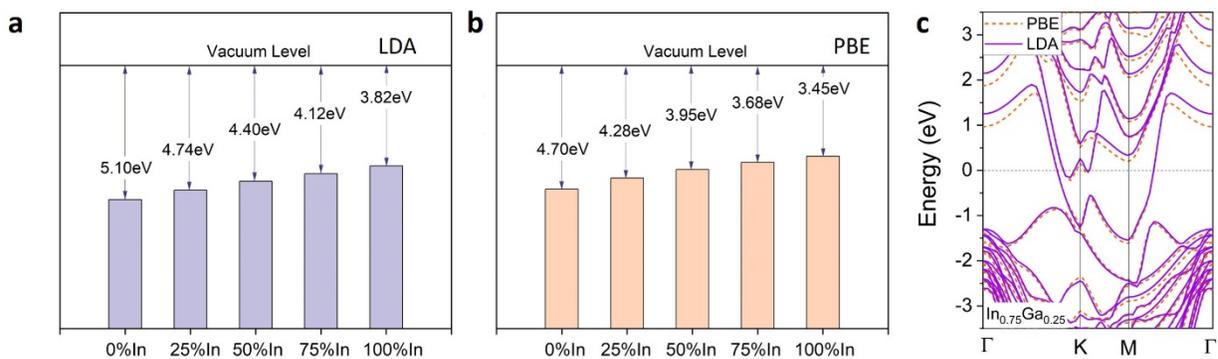

**Figure S11. Work functions and band structures computed using LDA and PBE exchange-correlation functionals.** Work functions of computed 2L In$_x$Ga$_{1-x}$/SiC systems (x=0, 0.25, 0.5, 0.75 and 1) computed by (a) LDA and (b) PBE exchange-correlation functionals. For comparison, the LDA and PBE work functions of free-standing 2L Gr are 4.57 and 4.26 eV, respectively, and the LDA and PBE work functions of graphene strained (by ~7%) to fit within the alloy/SiC supercells are 5.12 and 4.83 eV, respectively. All structures are relaxed using the corresponding functional except for 2L Gr which is relaxed using LDA only. (c) Band structure computed for 2L In$_{0.75}$Ga$_{0.25}$/SiC using LDA (solid purple line) and PBE (dashed orange line). The Fermi level is set at 0 eV. The VCA is used for alloy systems.

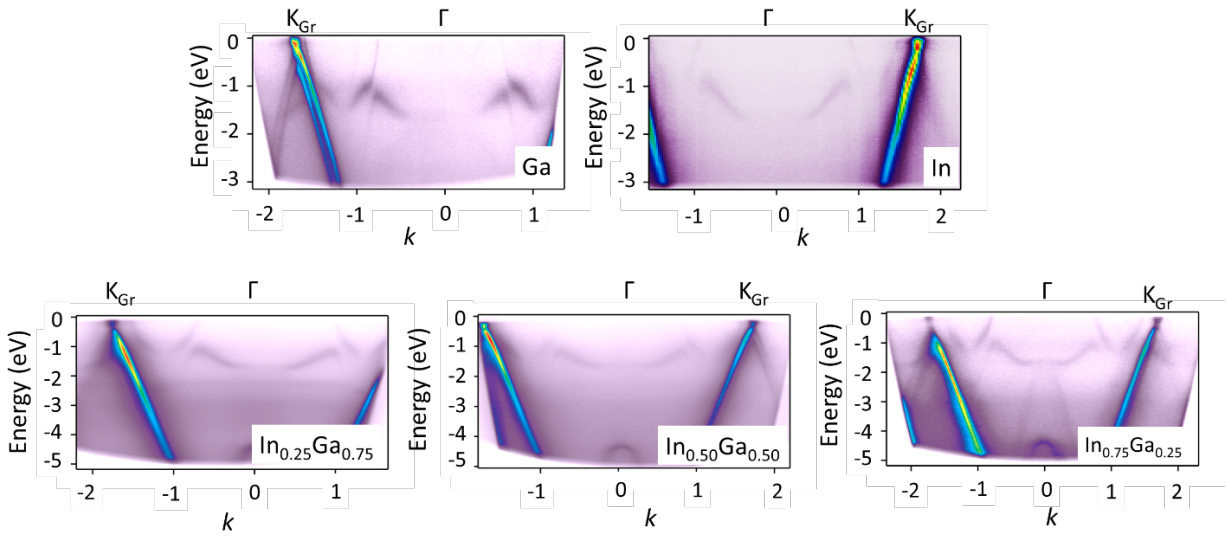

**Figure S12. ARPES-measured band structures of 2D In$_x$Ga$_{1-x}$/SiC systems along Γ-K$_{Gr}$.** The bright blue slopes correspond to the band of graphene. Dirac points are slightly below the Fermi level in all systems indicating that electrons are transferred to graphene from the pure elements and alloys.

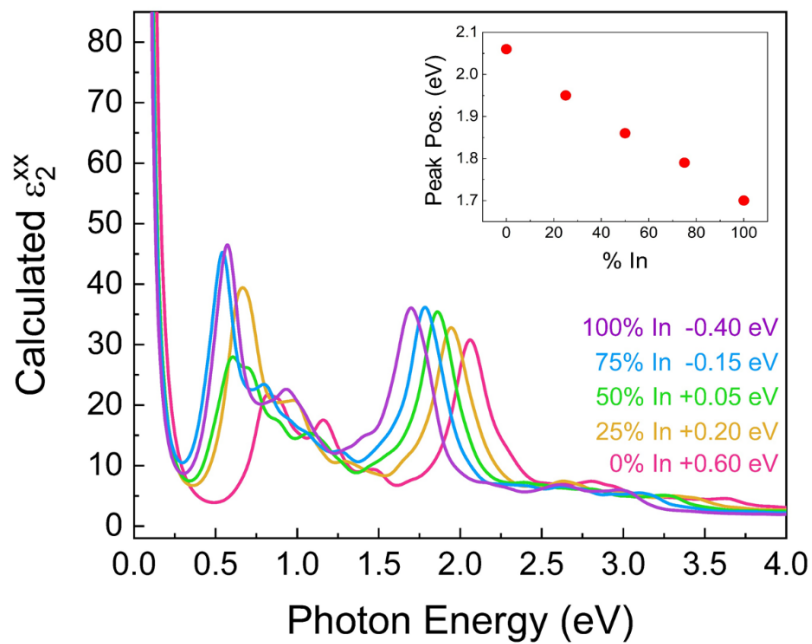

**Figure S13. Computed $\varepsilon_2^{xx}(E)$ with Fermi level shifted to match the ARPES results.** The inset indicates the major interband optical transition peak as a function of Indium content. The peak positions in the range ~1.5-2.0 eV are the same as in Figure 5d where the DFT Fermi level is used.

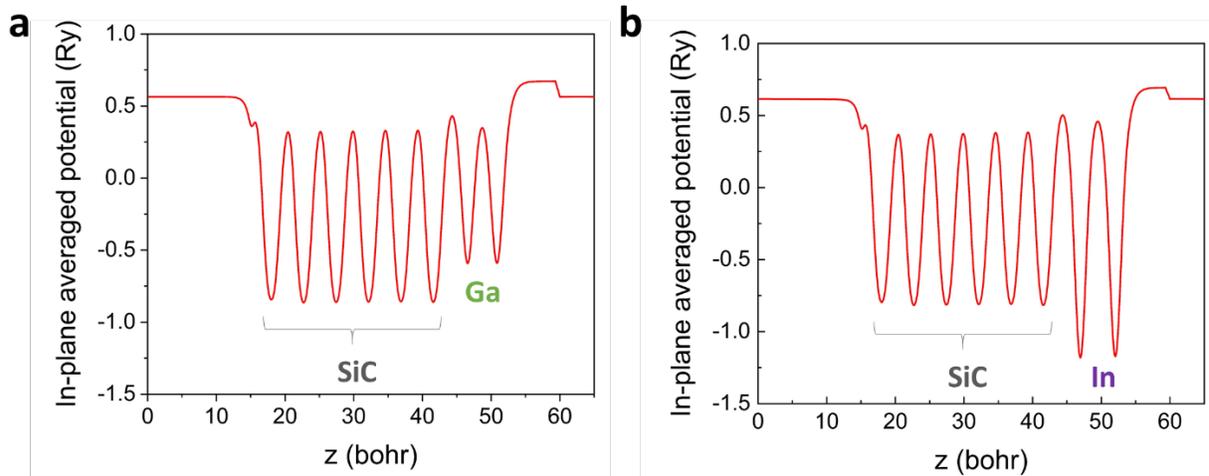

**Figure S14. In-plane averaged electrostatic potential.** In-plane averaged electrostatic potential as a function of distance *z* from bottom to top of the cell for (a) 2L Ga/SiC system and (b) 2L In/SiC system, demonstrating the potential well for In is deeper than that for Ga.

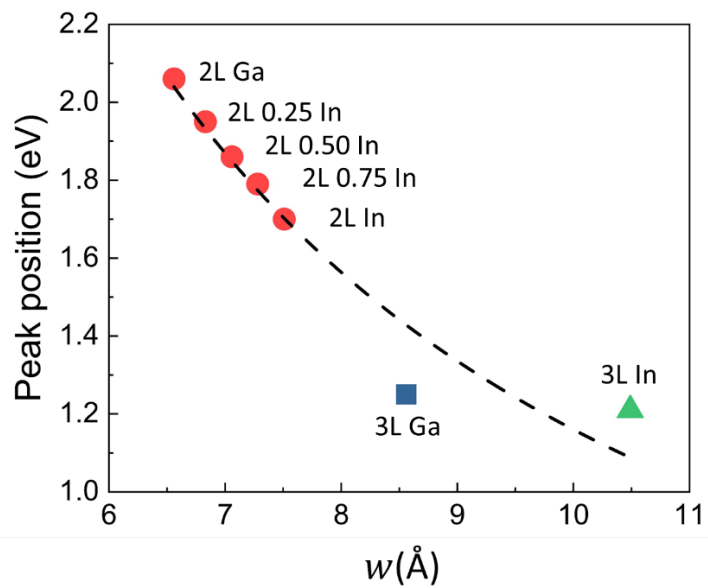

**Figure S15. Dependence of $\varepsilon_2$ peak position on effective quantum well thickness $w$.** The peak position of the dominant optical transition peak between ~1.0 to ~2.0 eV is plotted against $w$, which is defined to be the distance between the topmost metal layer and the interface Si layer together with the atomic radius of the metal element. The data for 3L Ga and 3L In is taken from Ref. [3]. The dashed line is the curve of best fit (peak position (eV) = $25.3w^{-1.3}$).

**Table S3. Structural parameters computed using LDA and PBE exchange-correlation functionals.** The interlayer distances $d_z$ between Si1 and M1, M1 and M2 as well as the SiC lattice constant $a_{SiC}$ are shown. Si1 is the Si layer adjacent to 2D metals. M1 and M2 is the bottom and top metal layer, respectively. The VCA is used for alloy systems.

| Unit: Å | | Ga | In$_{0.25}$Ga$_{0.75}$ | In$_{0.5}$Ga$_{0.5}$ | In$_{0.75}$Ga$_{0.25}$ | In | All |
|---|---|---|---|---|---|---|---|
| | $d_z$(Si1-M1) | 2.42 | 2.48 | 2.54 | 2.58 | 2.62 | |
| LDA | $d_z$(M1-M2) | 2.27 | 2.40 | 2.49 | 2.58 | 2.69 | |
| | $a_{SiC}$ | | | | | | 3.06 |
| | $d_z$(Si1-M1) | 2.47 | 2.54 | 2.59 | 2.64 | 2.68 | |
| PBE | $d_z$(M1-M2) | 2.38 | 2.49 | 2.59 | 2.68 | 2.77 | |
| | $a_{SiC}$ | | | | | | 3.09 |